\documentclass[aps,prd,reprint,groupedaddress,longbibliography]{revtex4-1}

\usepackage{graphicx}
\usepackage{amsmath}
\usepackage[%
  colorlinks=true,
  urlcolor=blue,
  linkcolor=blue,
  citecolor=blue
]{hyperref}

\usepackage{ulem}

\usepackage{calligra}
\usepackage{calrsfs}
\DeclareMathAlphabet{\mathcalligra}{T1}{calligra}{m}{n}
\DeclareMathAlphabet{\pazocal}{OMS}{zplm}{m}{n}

\usepackage{lipsum}

\begin{document}

\title{Probing feebly interacting dark matter with monojet searches}%

\author{J\'er\^ome Claude}%
\email{jerome.claude@carleton.ca}
\author{Ma\'ira Dutra}%
\email{mdutra@physics.carleton.ca}
\author{Stephen Godfrey}%
\email{godfrey@physics.carleton.ca}
\affiliation{Ottawa-Carleton Institute for Physics,
Carleton University, 1125 Colonel By Drive, Ottawa, Ontario K1S 5B6, Canada}

\date{\today}%

\begin{abstract}
Dark matter may consist of feebly interacting massive particles (FIMPs) that never thermalized with the cosmic plasma. Their relic density is achieved via freeze-in for a wide range of masses, significantly expanding the model space that can be tested compared to other production mechanisms. However, testing the tiny couplings required by freeze-in is challenging. We show that FIMPs can be probed by LHC searches for new physics in monojet events with large missing energy. We study a $Z'$ portal model in which gluon annihilation produces FIMPs in the early universe and today at colliders. Monojet searches by LHC Run 3 might discover new physics accounted for by FIMPs with mass in the MeV-TeV range.
\end{abstract}

\maketitle

\section{Introduction} \label{sec:intro}
The existence of dark matter (DM) particles is well-established from their gravitational interaction with ordinary matter~\cite{Bertone:2004pz,Bergstrom:2012fi,Gelmini:2015zpa}. However, their nature remains a major open problem. There are many ways to achieve the observed DM relic abundance ($\Omega h^2 \approx 0.120 \pm 0.001$~\cite{Planck:2018vyg}). The most studied approach is via freeze-out, in which weakly interacting massive particles (WIMPs) are initially thermalized with the 
standard model (SM) particles. Viable WIMP candidates have electroweak scale couplings and masses typically in the few GeV to few TeV range. While simplified freeze-out models are mostly in tension with experimental bounds, WIMPs are still well-motivated DM candidates~\cite{Baer:2014eja,Arcadi:2017kky,Roszkowski:2017nbc,Han:2022ubw}. 

Another possibility for the origin of dark matter that has been gaining the community's interest is the freeze-in mechanism, in which the SM bath produces feebly interacting massive particles (FIMPs) in out-of-equilibrium processes~\cite{Hall:2009bx,Bernal:2017kxu}. An appealing feature of freeze-in is the possibility for viable DM candidates with masses from a few keV~\cite{Chang:2019xva,Dvorkin:2020xga,Coskuner:2021qxo} up to the scale of the maximal temperature of the universe~\cite{Chung:1998ua,Carney:2022gse}. 

Despite the fact that the 
SM-FIMP couplings must be feeble enough for freeze-in to work, FIMP phenomenology is possible. In extra $U(1)$ extensions of the SM in which new gauge bosons interact feebly with SM fermions, freeze-in is testable by direct detection~\cite{Essig:2011nj,Chu:2011be,Essig:2015cda,Hambye:2018dpi,Heeba:2019jho,An:2020tcg}, electron beam dump, neutrino-electron scattering~\cite{Heeba:2019jho}, atomic parity violation, and collider experiments~\cite{Cosme:2021baj}. In models with mediators effectively coupled to photons, FIMP self-annihilation can generate keV X-ray lines~\cite{Brdar:2017wgy,Biswas:2019iqm}. Moreover, an early matter-dominated era increases the testability of any FIMP model~\cite{Bernal:2018ins,Cosme:2020mck}. 

Dark matter can be produced at colliders in association with detectable SM debris such as jets, leptons, and photons~\cite{Kahlhoefer:2017dnp,Trevisani:2018psx}. In particular, monojet signatures have come to be at the forefront of DM searches, as production of a single jet alongside dark matter is a common feature of DM models~\cite{Diehl:1994ff,Goodman:2010ku,Fox:2011pm,ATLAS:2021kxv,CMS:2021far}. Collider bounds on DM couplings and mass have the advantages of not depending on astrophysical uncertainties and being sensitive to very light DM candidates. The search for dark matter at colliders has focused 
on WIMPs due to their sizable couplings, but
searches for FIMPs have garnered more attention over the last decade. The collider bounds on FIMPs rely on the macroscopic decay of long-lived particles which usually play a role in the freeze-in process~\cite{Co:2015pka,Hessler:2016kwm,Calibbi:2018fqf,Belanger:2018sti,No:2019gvl,Okada:2020cue}. 

In this work, we show that freeze-in can also be tested by monojet searches at hadron colliders that arise from the missing transverse momentum of the DM particles. In a model where the SM fermions are neutral under an extra $U(1)$ gauge symmetry, FIMPs can be produced at the Large Hadron Collider (LHC) from a promptly decaying $Z'$ in association with a monojet signal. 

The paper begins in  Section~\ref{sec:model} with a description of the gluophilic $Z'$ model.  In Section~\ref{sec:relic} we describe how the dark matter relic abundance is calculated.  We give approximate expressions to help gain insights into freeze-in production although our results are found by numerically solving the full expressions. In Section~\ref{sec:outof} we examine the out-of-equilibrium condition for the $Z'$ which puts limits on the allowed parameter space.  Our results and discussion are given in Section~\ref{sec:results} where we include the various constraints on the parameter space including theoretical constraints, indirect detection limits based on Fermi-LAT searches, and the  constraints from monojet searches at the LHC.  Our conclusions are given in Section~\ref{sec:conclusions}.

\section{The gluophilic $Z'$ portal model}\label{sec:model}

Extending the SM by a new $U(1)$ gauge group, $U(1)'$, is one of the simplest frameworks for physics beyond the standard model (BSM) to address dark matter. An interesting possibility for a viable yet minimal DM portal model is that BSM states, including a DM candidate, are charged under $U(1)'$ while all SM states are neutral. In this case, an effective portal between the visible and dark sectors arises when BSM fermions carry SM hypercharge~\cite{Antoniadis:2009ze,Dudas:2009uq,Dudas:2012pb,Arcadi:2017jqd} and/or color~\cite{Dudas:2013sia,Ducu:2015fda,Bhattacharyya:2018evo}. 

We consider a model which includes a fermionic DM candidate $\chi$ and a set of BSM fermions $\Psi_i$ that are charged under $U(1)^\prime$ and are electroweak singlets. If the $\Psi_i$ are also charged under $SU(3)_c$ and are heavy enough, the new gauge boson $Z'$ will only interact with gluons via a dimension-6 operator. Unlike in simplified $Z'$ portal models, our $Z^\prime$ does not interact with SM fermions directly. The relevant Lagrangian for this gluophilic $Z'$ portal model is given by~\cite{Dudas:2013sia}
\begin{equation}\label{Eq:LagEff}
\begin{split}
    \pazocal{L} \supset 
    &~\frac{g_{Z^\prime}}{2}\bar \chi \gamma^\mu (1 - \gamma_5)\chi Z'_\mu\\
    &-\frac{\epsilon^{\mu\nu\rho\sigma}}{\Lambda^2} \Big( \partial^\alpha Z^\prime_\alpha Tr[G_{\mu\nu}G_{\rho\sigma}] - 2 Z^\prime_\mu Tr[G_{\nu\lambda} \partial^\lambda G_{\rho\sigma}]\Big) \,,
\end{split}\end{equation}
where $g_{Z'}$ is the gauge coupling associated with $U(1)'$, $\Lambda$ is the cut-off scale of the theory, and $G_{\mu\nu}$ is the gluon field strength.

Values of $\Lambda$ at intermediate scales are well-motivated by GUT theories and provide an elegant explanation for small FIMP couplings~\cite{Bhattacharyya:2018evo}. However, in order to study the testability of the model, we consider $\Lambda$ to be as small as possible for freeze-in to work. Assuming there are $N_\Psi$ nearly degenerate $\Psi_i$ fermions with mass $M$ and dark chiral charges satisfying $q_{\Psi^i_L}-q_{\Psi^i_R} = 1$, $\Lambda$ is related to the parameters of the UV complete theory by 
\begin{equation}\label{Eq:Lambda}
\Lambda \simeq \sqrt{\frac{24\pi}{N_\Psi \alpha_s g_{Z^\prime}}} M\,,
\end{equation}
where $\alpha_s$ is the strong coupling constant evaluated at the $Z$ mass. The actual value of $N_\Psi$ is not relevant in our effective analysis, as it only re-scales the $Z'$ coupling to gluons. The only limit that we are aware of for an electroweak singlet color triplet Dirac fermion is derived from the running of $\alpha_s$ in high energy collisions at the LHC~\cite{Llorente:2018wup}. Ref.~\cite{Llorente:2018wup} finds $M\gtrsim {\cal O}(100)$~GeV for three degenerate fermions, although that limit can be off by a factor of 2 due to theoretical uncertainties. This merits future study which lies outside the scope of the present work. 

In order to make explicit the implications of our results, in what follows we treat $M$ as a free parameter. The other free parameters in our analysis are the dark matter and $Z'$ masses, $m_\chi$ and $m_{Z'}$, the number of heavy fermions $N_\Psi$, and the $U(1)'$ gauge coupling $g_{Z'}$. Although we parametrize this model in terms of quantities that are physically meaningful for the UV theory, we present results at scales where the effective model remains valid.

An important feature of this model is the momentum dependence of the interaction between gluons and the $Z'$. For $Z'_\lambda$, $G_\mu$, and $G_\nu$ with four-momenta $k$, $p_1$, and $p_2$, respectively, the $Z' gg$ vertex is given by
\begin{equation}\label{Eq:ZpGGvertex}
    \frac{N_\Psi \alpha_s g_{Z'}}{12\pi M^2} \delta^{ab} (A_{Z'}^{\lambda \mu \nu} + A_g^{\lambda \mu \nu})\,,
\end{equation}
with the momentum-dependent terms $A_{Z'}^{\lambda \mu \nu}$ and $A_g^{\lambda \mu \nu}$ defined as
\begin{align}
&A_{Z'}^{\lambda \mu \nu} \equiv -2 k^\lambda \epsilon^{\mu \nu \rho \sigma} {p_1}_\rho {p_2}_\sigma \nonumber\\
&A_g^{\lambda \mu \nu} \equiv (p_1 \cdot p_2) \epsilon^{\mu \nu \lambda \alpha} (p_1-p_2)_\alpha \nonumber\\
&\hspace{1.2cm} + {p_1}_\rho {p_2}_\sigma ({p_2}^\mu \epsilon^{\nu \lambda \rho \sigma} - {p_1}^\nu \epsilon^{\mu \lambda \rho \sigma})\nonumber\,.
\end{align}
The contraction between one of the momenta and the corresponding polarization vector is exactly zero for an on-shell particle. As a consequence, in processes involving the vertex $Z'gg$, at least one of the three gauge bosons must be off-shell. 

\section{Freeze-in production of dark matter}\label{sec:relic}

In this model, dark matter can be produced in the early universe from gluon self-annihilation via the s-channel exchange of $Z'$ bosons. The reduced cross-section for this process reads
\begin{align}
\hat \sigma_{gg \to \bar \chi \chi} &= \frac{2 N_\Psi^2 \alpha_s^2 g_{Z'}^4}{32\pi^3} \frac{m_\chi^2}{m_{Z'}^4 M^4} \frac{ s^3 (s-m_{Z'}^2)^2 \sqrt{1-\frac{4m_\chi^2}{s}}}{(s-m_{Z'}^2)^2+m_{Z'}^2 \Gamma_{Z'}^2} \nonumber\\
&\simeq \frac{2 N_\Psi^2 \alpha_s^2 g_{Z'}^4}{32\pi^3} \frac{m_\chi^2 s^3}{m_{Z'}^4 M^4} \sqrt{1-\frac{4m_\chi^2}{s}} \,,
\end{align}
where $\sqrt{s}$ is the center-of-mass energy, $\Gamma_{Z'}$ the total decay width of $Z'$, and the approximation holds for $\Gamma_{Z'}^2 \ll m_{Z'}^2$. Close to the $Z'$ pole, this cross-section is suppressed instead of enhanced because the $Z'$ must be off-shell~\cite{Bhattacharyya:2018evo}. It is worth mentioning that only the axial coupling between $\chi$ and $Z'$ contributes to this process. In the limit $\Gamma_{Z'}^2 \ll m_{Z'}^2$ and $s \gg 4m_\chi^2$, the reaction rate density for the $gg \to \bar \chi \chi$ process is given by
\begin{equation}
\gamma_{gg\to \bar \chi \chi}(T) \approx 6.4\times 10^{-3} N_\Psi^2 \alpha_s^2 g_{Z'}^4 \frac{m_\chi^2 T^{10}}{m_{Z'}^4 M^4} e^{-\frac{m_\chi}{T}}\,.
\end{equation}
In this expression, we have approximated the effect of the Boltzmann suppression on the dark matter number density when $T<m_\chi$ by an exponential decrease.

The freeze-in production takes place when dark matter is initially absent and is produced from thermal bath species in out-of-equilibrium processes. The high temperature dependence of this reaction rate means that the freeze-in process starts to be effective as soon as a thermal bath is established, at the maximal temperature of the universe $T_\text{\tiny{MAX}}$. We can determine the region of our parameter space in which the out-of-equilibrium condition is satisfied, and thus the freeze-in regime holds, by requiring that the reaction rate $\gamma_{gg \to \bar \chi \chi}(T)/n_g(T)$, with $n_g$ the gluon number density, be smaller than the Hubble rate $H(T)$ at $T = T_\text{\tiny{MAX}}$:
\begin{equation}
    0.013 N_\Psi^2 \alpha_s^2 g_{Z'}^4 \frac{\sqrt{g_e(T_\text{\tiny{RH}})}}{g_e(T_\text{\tiny{MAX}})}\frac{m_\chi^2 T_\text{\tiny{RH}}^2 T_\text{\tiny{MAX}}^3 M_{Pl}}{M^4 m_{Z'}^4} e^{-\frac{m_\chi}{T_\text{\tiny{MAX}}}}
\lesssim 1 \,,
\label{eq:FIboundary}
\end{equation}
where $T_\text{\tiny{RH}}$ is the inflationary reheat temperature, $g_e(T)$ and $g_s(T)$ are respectively the energetic and entropic relativistic degrees of freedom at a given temperature, and $M_{Pl}\approx 2.43 \times 10^{18}$~GeV is the reduced Planck mass.

In order to evaluate the relic abundance of dark matter in this scenario, we numerically solve a coupled set of Boltzmann fluid equations for the number density of $\chi$ and the energy densities of radiation and of the inflaton field~\cite{Bhattacharyya:2018evo}. However, as explained in detail in Ref.~\cite{Dutra:2019gqz}, one can find an approximate expression for the relic density which takes into account freeze-in during reheating. In the present case, we find
\begin{widetext}
\begin{equation}
\frac{\Omega h^2}{0.12} \approx \left(\frac{76}{g_\text{eff}}\right)^{3/2} \left(\frac{N_\Psi}{3}\right)^2 \left(\frac{g_{Z'}}{0.01}\right)^4 \left(\frac{m_\chi}{0.6 \text{GeV}}\right)^3 \left(\frac{2 \text{TeV}}{M}\right)^4 \left(\frac{1 \text{GeV}}{m_{Z'}}\right)^4 \left(\frac{T_\text{\tiny{RH}}}{1\text{GeV}}\right)^5 \left[1+8.13 \,r \left(\frac{T_\text{\tiny{RH}}}{1GeV}\right)^2 \left(\frac{0.6 \text{GeV}}{m_\chi}\right)^2\right]\,.
 \label{eq:FIrelic}
\end{equation}
\end{widetext}

In this approximation, we assumed that the relativistic degrees of freedom do not change significantly during freeze-in ($g_e \approx g_s \equiv g_\text{eff}$). The first term accounts for production during the radiation era ($T<T_\text{\tiny{RH}}$), where we have assumed $m_\chi \ll T_\text{\tiny{RH}}$. The second term accounts for production during reheating, and we have assumed $T_\text{\tiny{RH}} \ll m_\chi \ll T_\text{\tiny{MAX}}$. We, therefore, parameterize the contribution of freeze-in during reheating with the variable $r$: $r=0$ ($r=1$) when $m_\chi<T_\text{\tiny{RH}}$ ($m_\chi>T_\text{\tiny{RH}}$).  We have also neglected the exponential suppression on the rate when $T \sim m_\chi$. Eq.~\ref{eq:FIrelic} will be useful to interpret our numerical results for the relic density of $\chi$.

\section{Condition for a nonthermal $Z'$} \label{sec:outof}

In the freeze-in analysis above, we have assumed that only gluons produce DM. However, if $Z'$ bosons interact strongly enough with gluons, they would be part of the cosmic thermal bath and thus contribute to the freeze-in of $\chi$. The leading process that would thermalize the $Z'$ and gluons is $q \bar q \to Z' g$, via gluon s-channel exchange. Although we compute the reaction rate numerically in our results, we found it to be well approximated by the following expression:
\begin{equation}
    \gamma_{q\bar q \to Z' g}(T) \approx 1.4 \times 10^{-3} N_\Psi^2 \alpha_s^3 g_{Z'}^2 \frac{T^{10}}{m_{Z'}^2 M^4} e^{-\frac{\text{max}(2m_q,m_{Z'})}{T}}\,.
\end{equation}
Since the thresholds for this process are $s>4m_q^2$ and $s>m_{Z'}^2$, we have approximated the effect of the thresholds by an exponential decrease as $T$ approaches the maximum between the $2m_q$ and $m_{Z'}$  where $m_q$ is the quark mass. In the results below, we will only show the part of the relic density contours satisfying the out-of-equilibrium condition for the $Z'$, namely $\gamma_{q\bar q \to Z' g}(T)/n_q(T)<H(T)$ at $T=T_\text{\tiny{MAX}}$.

Finally, we note that the sub-leading processes are two-loop suppressed. Processes with two $Z'gg$ vertices such as $gg\to Z' Z'$ are suppressed by a factor of $1/\Lambda^8$. There are also processes featuring both the $Z'gg$ vertex and QCD loops. Among those, we expect $Z'\to q\bar{q}$ to be the leading one, and we found it to be negligible.

\section{Results and discussion}\label{sec:results}

We are now in a position to present the constraints on the viable parameter space of feebly interacting dark matter in the context of the gluophilic $Z'$ portal model discussed above. 

We start with the theoretical constraints that ensure the consistency of the model. Since we are working in the effective regime, the scale of new physics, $\Lambda$, must be above all relevant energy scales. This condition is satisfied by the allowed values of $M$ via Eq. \ref{Eq:Lambda}. Additionally, the presence of axial couplings lead to the violation of unitarity of the process $\bar \chi \chi \to \bar \chi \chi$ at high energies~\cite{Kahlhoefer:2015bea}. As a consequence, dark matter cannot be arbitrarily heavier than $Z'$:
\begin{equation}
    m_\chi \lesssim \sqrt{2\pi}\frac{m_{Z'}}{g_{Z'}}\,.
\end{equation}

\begin{figure}
    \includegraphics[width=.95\linewidth]{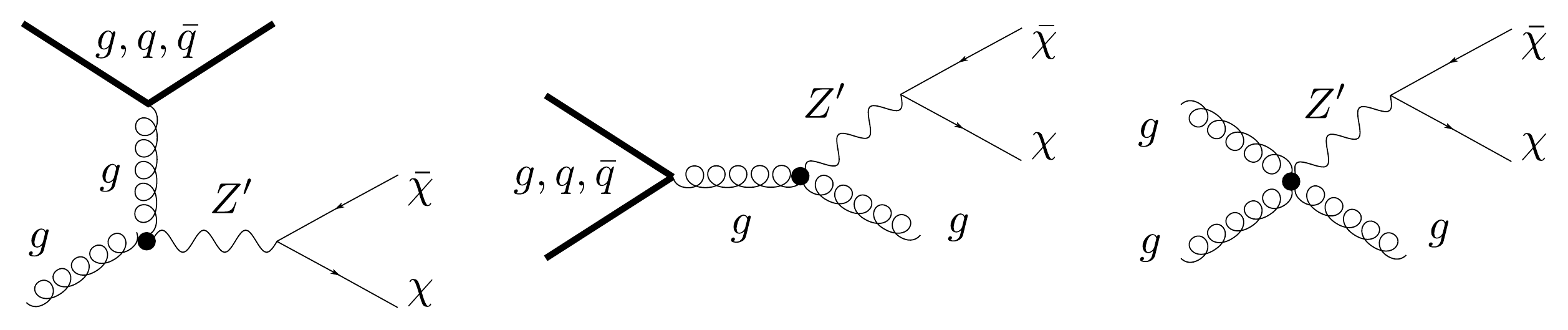}
    \caption{Processes leading to monojet signals. The thick lines represent gluons ($g$) and quarks ($q = u$, $d$, $c$, and $s$).
    \label{fig:monojetdiag}}
\end{figure}

We now consider the experimental constraints on our model. Proton-proton collisions at the LHC can produce $Z'$ bosons which promptly decay into dark matter. The Feynman diagrams allowing for monojet signals are depicted in Fig.~\ref{fig:monojetdiag}. The ATLAS collaboration searched for new physics in events with an energetic jet and large missing transverse energy ($E_T^\text{miss}$) using 139~fb$^{-1}$ of data collected at $\sqrt{s}=13$~TeV~\cite{ATLAS:2021kxv}. Since no excess over the SM background was found, they provided model-independent upper limits on the visible cross section for different $E_T^\text{miss}$ event selection criteria. In this work, we use the limit with the strictest selection criteria to optimize signal-to-background ratio and, in turn, sensitivity. For $E_T^\text{miss}>1200$~GeV, values of $\sigma \times A \times \epsilon$ above 0.3~fb are excluded at 95\% confidence level. Using {\tt MadGraph5\_aMC\@NLO}~\cite{Alwall:2014hca}, we generated events featuring a single jet with transverse momentum $p_T > 1200$~GeV and pseudorapidity $|\eta|<2.4$, which accounts for the acceptance. We assume $\epsilon \sim 1$ and found that reducing the efficiency to as low as 10\% does not significantly change the limits. Because parton-level constraints must affect partons individually, the $E_T^\text{miss}$ constraint  was implemented as a constraint on the jet $p_T$  by conservation of momentum between the outgoing jet and the invisible $\chi\bar{\chi}$ system.

Dwarf Spheroidal Galaxies (dSphs) are ideal targets for indirect detection experiments to search for signals of dark matter annihilation or decay~\cite{Strigari:2018utn}. In the present model, dark matter annihilates exclusively into gluons. After showering and hadronization, the gluons generate a continuum flux of gamma rays ~\cite{Cirelli:2010xx}. We computed the annihilation cross section $\langle \sigma v\rangle$ for the process $\chi \bar \chi \to gg$ using {\tt MadDM}~\cite{Ambrogi:2018jqj}. Within the {\tt MadDM} framework, we extracted the exclusion limits on $\langle \sigma v\rangle$ at 95\% confidence level based on the Fermi-LAT searches for gamma rays coming from DM annihilation in dSphs.

Given the absence of a tree-level scattering of $\chi$ off of quarks, direct detection in the class of models considered in this work is hardly possible. In our specific case of a gluophilic $Z'$, the DM-nuclei matrix element vanishes \cite{Dudas:2013sia}.

\begin{figure}
    \includegraphics[width=\linewidth]{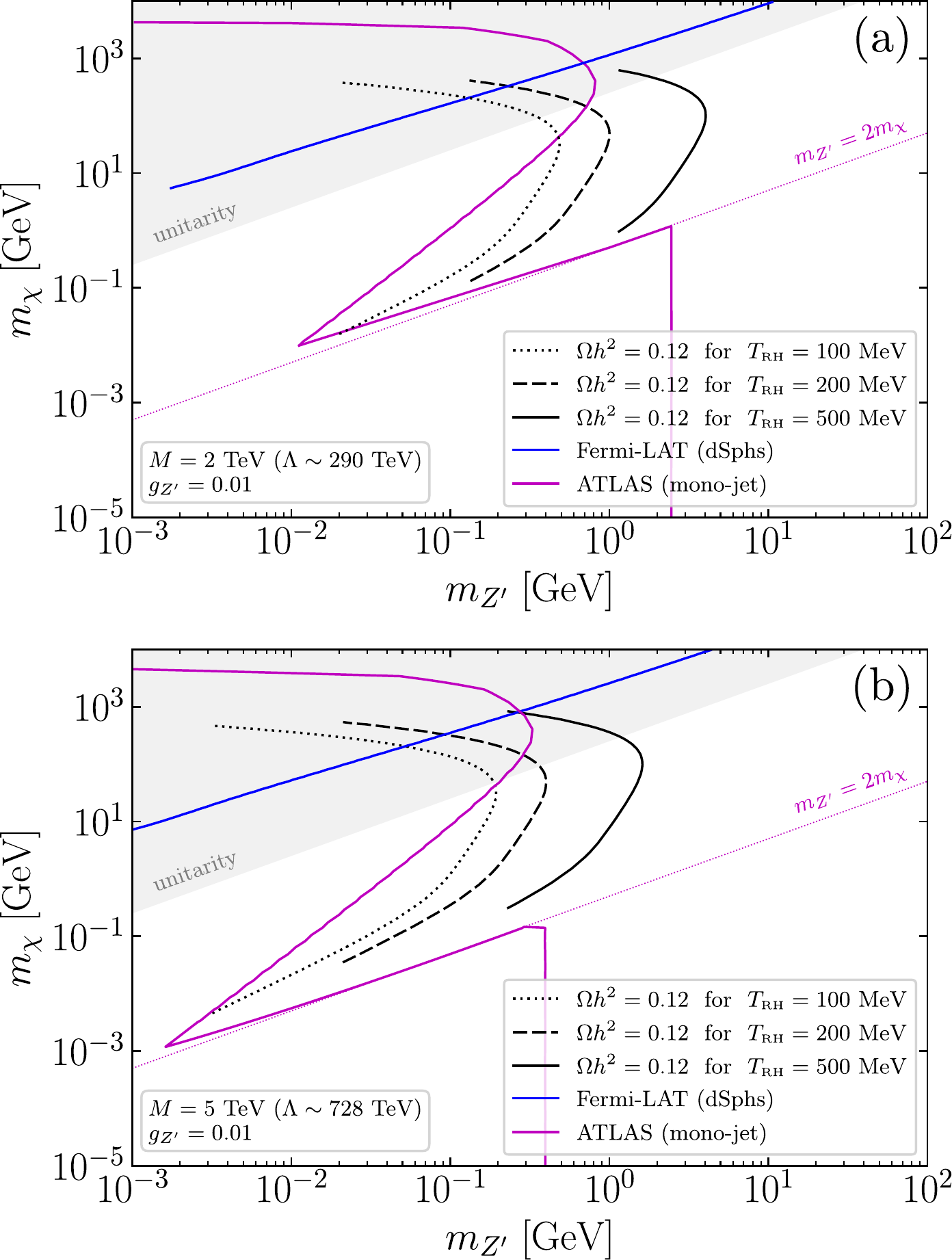}
    \caption{Contours for the correct relic density of our dark matter candidate $\chi$ in the $(m_{Z'},m_\chi)$ parameter plane, for three values of reheat temperature (black contours). Gray regions are excluded by the breakdown of perturbative unitarity. Regions to the left of the magenta curves are excluded by ATLAS monojet searches and regions above the blue lines are excluded by Fermi-LAT searches for gamma rays from DM self-annihilation in the Milky Way's dwarf spheroidal galaxies. The Fermi-LAT limits are within the region excluded by the unitarity bound, being therefore invalid. Monojet searches can test regions of our parameter space in which DM is produced via freeze-in, for DM masses spanning from the MeV up to the TeV scale. \label{fig:mZp-mdm}}
\end{figure}

\begin{figure}
    \includegraphics[width=\linewidth]{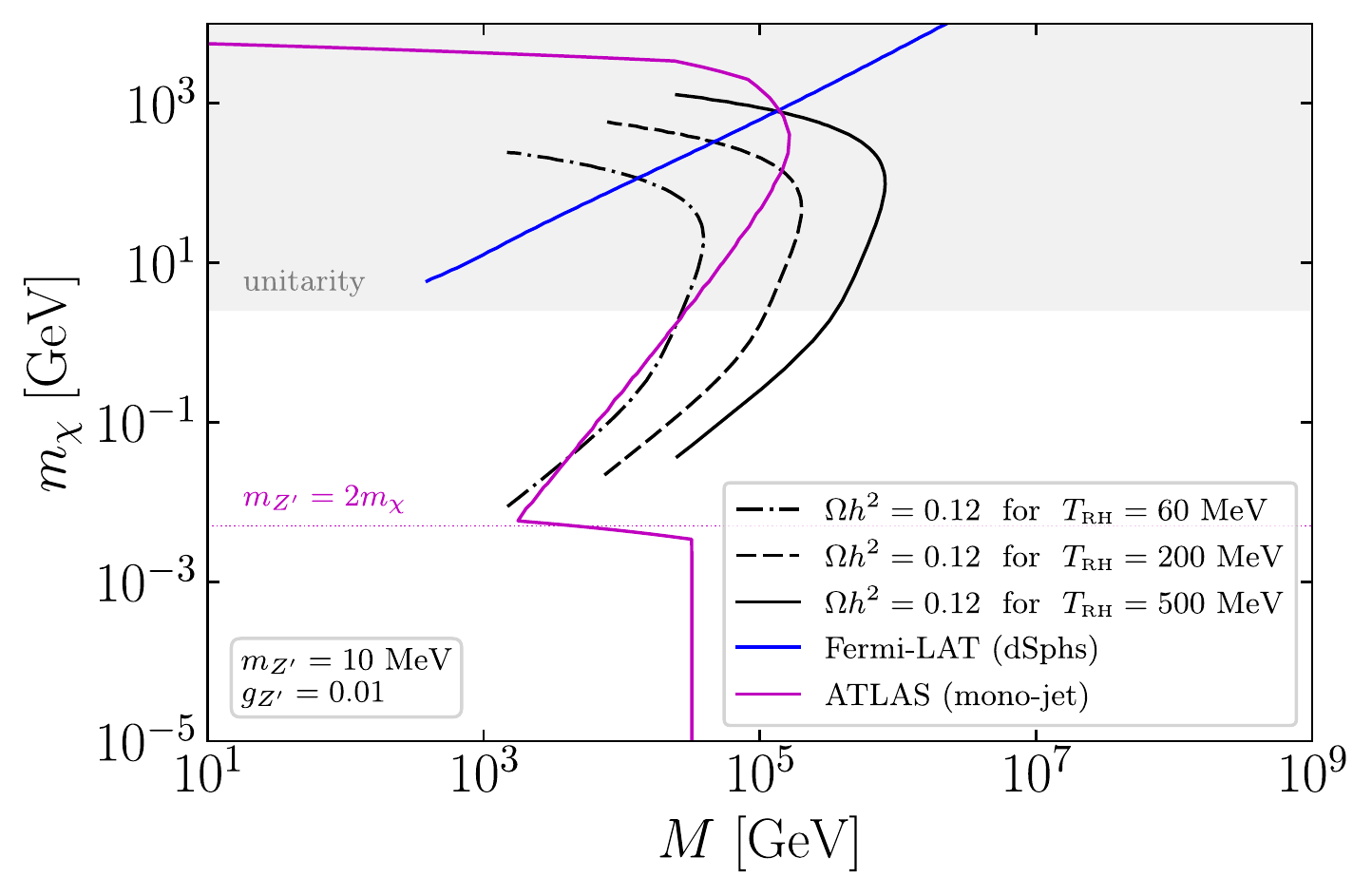}
    \caption{Contours for the correct relic density of our dark matter candidate $\chi$ in the $(M,m_\chi)$ parameter plane, for three values of reheat temperature (black contours). Same color labeling as in Fig.~\ref{fig:mZp-mdm}. Monojet searches can test freeze-in even for the case of an MeV-scale gluophilic $Z'$. \label{fig:M-mdm}}
\end{figure}

Our results are summarized in Figs. \ref{fig:mZp-mdm} and \ref{fig:M-mdm}, where we show the constraints on the parameter space providing the observed
dark matter relic density (black contours) in the parameter planes $m_{Z^\prime}-m_\chi$ and $M-m_\chi$, respectively. The gray regions are excluded by the theoretical bounds discussed above, namely perturbative unitarity. Regions where the breakdown of the effective approach occurred are at higher values of masses than is shown in the figures. The regions to the left of the magenta contours are excluded by ATLAS monojet searches~\cite{ATLAS:2021kxv}, whereas the regions above the blue contours are excluded by the Fermi-LAT searches for gamma rays from DM annihilation in dSphs~\cite{Ambrogi:2018jqj}.

As discussed above, the relic density depends strongly on the reheat temperature $T_\text{\tiny{RH}}$. $T_\text{\tiny{RH}}$ is restricted by cosmological observations to be above the MeV scale~\cite{Hannestad:2004px,Jedamzik:2006xz,Kawasaki:2017bqm}. We treat $T_\text{\tiny{RH}}$ and $T_\text{\tiny{MAX}}$ as free parameters. The ratio $T_\text{\tiny{MAX}}/T_\text{\tiny{RH}} \sim \frac{(M_{Pl} H_I)^{1/4}}{\sqrt{T_\text{\tiny{RH}}}}$, with $M_{Pl}$ the Planck mass and $H_I$ the Hubble rate at the end of inflation, may be large ~\cite{Giudice:2000ex}. We set $T_\text{\tiny{MAX}}/T_\text{\tiny{RH}}=100$ for illustrative purposes, but our conclusions are not significantly affected by this choice. Larger values of this ratio allow for heavier DM, as the viable region of our parameter space in which $T_\text{\tiny{RH}} < m_\chi < T_\text{\tiny{MAX}}$ would be larger. In Figs. \ref{fig:mZp-mdm} and \ref{fig:M-mdm}, we consider four  representative values for $T_\text{\tiny{RH}}$: $60$~MeV (dot-dashed black contours), $100$~MeV (dotted black contours), $200$~MeV (dashed black contours), and $500$~MeV (solid black contours). All contours shown for the correct relic density satisfy the out-of-equilibrium conditions under which the freeze-in regime holds (Eq.~\ref{eq:FIboundary}) and the $Z'$ is not part of the thermal bath. By ensuring that Eq. \ref{eq:FIboundary} is satisfied, we find that our FIMP candidate cannot be too light. Moreover, we find that low enough values of $m_{Z'}$ \textit{and} high enough values of $T_\text{\tiny{MAX}}$ bring the $Z'$ into equilibrium with gluons. Therefore, the viable region of our parameter space where the freeze-in is testable and driven by gluon annihilation relies on low-scale reheating.

With Eq.~\ref{eq:FIrelic}, one can easily understand the features of the relic density contours and how they would change with different values of the free parameters. The first term of Eq.~\ref{eq:FIrelic} is a good approximation when $m_\chi < T_\text{\tiny{RH}}$, whereas the second term is a good approximation when $m_\chi > T_\text{\tiny{RH}}$. However, when $m_\chi > T_\text{\tiny{MAX}}$, one can see the Boltzmann suppression of the relic density, which is not considered by our approximation. To compensate for this suppression, light mediators $Z'$ and $\Psi_i$ are needed. Because of the high temperature dependence of the production rate, for a given value of $m_\chi$, the higher the reheat scale, the more dark matter is produced. Therefore, the heavier the mediators must be to produce the same amount of $\chi$.

The theoretical bounds set strong upper limits on the dark matter mass. In the upper panel of Fig.~\ref{fig:mZp-mdm}, we set $M = 2$~TeV, corresponding to $\Lambda \approx 291.4$~TeV. We observe that the unitarity bound excludes the limits from Fermi-LAT. Lower values of $M$ would enhance the Fermi-LAT limit and make it complementary to the ATLAS limit. However, such low values of $M$ cause the $Z'$ to thermalize with gluons. Higher values of $M$ are nevertheless of phenomenological interest. In the lower panel of Fig.~\ref{fig:mZp-mdm}, we set $M = 5$~TeV ($\Lambda \approx 728.5$~TeV). In this case, the process $gg \to \bar \chi \chi$ is suppressed with respect to the upper panel case, and lower values of $m_{Z'}$ become viable.

For a given value of $M$, the lighter the $Z'$, the stronger the indirect detection and monojet cross sections. However, sub-GeV $Z'$ bosons are also of interest in our context. In Fig.~\ref{fig:M-mdm}, we set $g_{Z'} = 0.01$ and $m_{Z'} = 10$~MeV. For such a light $Z'$ boson, the upper limit on $m_\chi$ coming from unitarity is much stronger, ruling out the region of our parameter space probed by indirect detection. Interestingly, we find that for reheat scales much lower than hundreds of MeV, our dark matter candidate is completely ruled out by the monojet bounds. For reheat scales above hundreds of MeV, our dark matter candidate completely evades the current monojet limits.

Our results show that monojet searches strongly constrain the mass spectrum of a gluophilic $Z'$ portal model. In the region below the magenta dotted lines (where $m_{Z'}=2m_\chi$), $Z'$ is resonantly produced and decays into dark matter. From Eq.~\ref{Eq:ZpGGvertex}, we note that this is allowed by the $A_g^{\lambda \mu \nu}$ contribution to the $Z'gg$ vertex. The production rate of DM at the LHC becomes insensitive to the DM mass when the mediator is too heavy. The value at which this occurs depends on the interaction strength $g_{Z'}/M^2$. On the other hand, the off-shell mediator region ($m_{Z'}<2m_\chi$) is strongly probed by monojet searches, unlike what we find in simplified $Z'$ portal models. At values of $m_\chi$ approaching $\sqrt{s}/2$, DM becomes kinematically inaccessible for monojet searches. Therefore, indirect detection bounds offer an opportunity to probe dark matter candidates too heavy to be produced at colliders.

Upcoming measurements by collider and indirect detection experiments will probe new viable regions of our parameter space. The recent launch of the LHC Run 3 and the planned High Luminosity runs will push both the energy and precision frontiers further than before~\cite{Baum:2017kfa,ATL-PHYS-PUB-2022-018}. Future observatories such as the Cherenkov Telescope Array (CTA) and the Southern Wide field-of-view Gamma-ray Observatory (SWGO) will reach unprecedented sensitivity to DM particles in the 100 GeV -- $10^6$~GeV mass range~\cite{Viana:2021smp}. Finally, it is worth mentioning that a reheating scale roughly below $150$~MeV might lead to detectable effects~\cite{Delos:2021rqs}.

\section{Conclusions}\label{sec:conclusions}

The freeze-in mechanism has become a popular explanation for the origin of dark matter. Frozen-in DM particles, the FIMPs, evade most of the current experimental bounds and motivate new searches for DM in a broad mass range. We have shown that monojet searches at the LHC are able to constrain the parameter space of a FIMP candidate. We considered a model in which gluons annihilate into FIMPs by exchanging a gluophilic $Z'$. Feeble DM interactions consistent with freeze-in are easily accomplished in this model, as the coupling between $Z'$ and gluons is loop-induced. The high momentum dependence of this effective coupling has a two-fold role in our results. On the one hand, it causes the freeze-in to occur during the post-inflationary reheating period. On the other hand, it leads to stringent monojet bounds, even in the off-shell mediator regime. A large region of our parameter space, for FIMP masses between the MeV scale and the collider threshold of $6.5$~TeV, is already excluded by monojet limits. The LHC and future MeV gamma-ray telescopes will further test this model in the coming years, placing the usually elusive FIMPs at the edge of detection.

\begin{acknowledgments}
We are grateful to Gopolang Mohlabeng and Olivier Mattelaer for their help with {\tt MadGraph5\_aMC\@NLO}.
This work was supported by the Natural Sciences and Engineering Research Council of Canada under grant No. SAPIN-
2016-00041.
\end{acknowledgments}


\begin{thebibliography}{61}%
\makeatletter
\providecommand \@ifxundefined [1]{%
 \@ifx{#1\undefined}
}%
\providecommand \@ifnum [1]{%
 \ifnum #1\expandafter \@firstoftwo
 \else \expandafter \@secondoftwo
 \fi
}%
\providecommand \@ifx [1]{%
 \ifx #1\expandafter \@firstoftwo
 \else \expandafter \@secondoftwo
 \fi
}%
\providecommand \natexlab [1]{#1}%
\providecommand \enquote  [1]{``#1''}%
\providecommand \bibnamefont  [1]{#1}%
\providecommand \bibfnamefont [1]{#1}%
\providecommand \citenamefont [1]{#1}%
\providecommand \href@noop [0]{\@secondoftwo}%
\providecommand \href [0]{\begingroup \@sanitize@url \@href}%
\providecommand \@href[1]{\@@startlink{#1}\@@href}%
\providecommand \@@href[1]{\endgroup#1\@@endlink}%
\providecommand \@sanitize@url [0]{\catcode `\\12\catcode `\$12\catcode
  `\&12\catcode `\#12\catcode `\^12\catcode `\_12\catcode `\%12\relax}%
\providecommand \@@startlink[1]{}%
\providecommand \@@endlink[0]{}%
\providecommand \url  [0]{\begingroup\@sanitize@url \@url }%
\providecommand \@url [1]{\endgroup\@href {#1}{\urlprefix }}%
\providecommand \urlprefix  [0]{URL }%
\providecommand \Eprint [0]{\href }%
\providecommand \doibase [0]{http://dx.doi.org/}%
\providecommand \selectlanguage [0]{\@gobble}%
\providecommand \bibinfo  [0]{\@secondoftwo}%
\providecommand \bibfield  [0]{\@secondoftwo}%
\providecommand \translation [1]{[#1]}%
\providecommand \BibitemOpen [0]{}%
\providecommand \bibitemStop [0]{}%
\providecommand \bibitemNoStop [0]{.\EOS\space}%
\providecommand \EOS [0]{\spacefactor3000\relax}%
\providecommand \BibitemShut  [1]{\csname bibitem#1\endcsname}%
\let\auto@bib@innerbib\@empty
\bibitem [{\citenamefont {Bertone}\ \emph {et~al.}(2005)\citenamefont
  {Bertone}, \citenamefont {Hooper},\ and\ \citenamefont
  {Silk}}]{Bertone:2004pz}%
  \BibitemOpen
  \bibfield  {author} {\bibinfo {author} {\bibfnamefont {Gianfranco}\
  \bibnamefont {Bertone}}, \bibinfo {author} {\bibfnamefont {Dan}\ \bibnamefont
  {Hooper}}, \ and\ \bibinfo {author} {\bibfnamefont {Joseph}\ \bibnamefont
  {Silk}},\ }\bibfield  {title} {\enquote {\bibinfo {title} {{Particle dark
  matter: Evidence, candidates and constraints}},}\ }\href {\doibase
  10.1016/j.physrep.2004.08.031} {\bibfield  {journal} {\bibinfo  {journal}
  {Phys. Rept.}\ }\textbf {\bibinfo {volume} {405}},\ \bibinfo {pages}
  {279--390} (\bibinfo {year} {2005})},\ \Eprint
  {http://arxiv.org/abs/hep-ph/0404175} {arXiv:hep-ph/0404175} \BibitemShut
  {NoStop}%
\bibitem [{\citenamefont {Bergstrom}(2012)}]{Bergstrom:2012fi}%
  \BibitemOpen
  \bibfield  {author} {\bibinfo {author} {\bibfnamefont {Lars}\ \bibnamefont
  {Bergstrom}},\ }\bibfield  {title} {\enquote {\bibinfo {title} {{Dark Matter
  Evidence, Particle Physics Candidates and Detection Methods}},}\ }\href
  {\doibase 10.1002/andp.201200116} {\bibfield  {journal} {\bibinfo  {journal}
  {Annalen Phys.}\ }\textbf {\bibinfo {volume} {524}},\ \bibinfo {pages}
  {479--496} (\bibinfo {year} {2012})},\ \Eprint
  {http://arxiv.org/abs/1205.4882} {arXiv:1205.4882 [astro-ph.HE]} \BibitemShut
  {NoStop}%
\bibitem [{\citenamefont {Gelmini}(2015)}]{Gelmini:2015zpa}%
  \BibitemOpen
  \bibfield  {author} {\bibinfo {author} {\bibfnamefont {Graciela~B.}\
  \bibnamefont {Gelmini}},\ }\bibfield  {title} {\enquote {\bibinfo {title}
  {{The Hunt for Dark Matter}},}\ \ }(\bibinfo {year} {2015})\ pp.\ \bibinfo
  {pages} {559--616},\ \Eprint {http://arxiv.org/abs/1502.01320}
  {arXiv:1502.01320 [hep-ph]} \BibitemShut {NoStop}%
\bibitem [{\citenamefont {Aghanim}\ \emph {et~al.}(2020)\citenamefont {Aghanim}
  \emph {et~al.}}]{Planck:2018vyg}%
  \BibitemOpen
  \bibfield  {author} {\bibinfo {author} {\bibfnamefont {N.}~\bibnamefont
  {Aghanim}} \emph {et~al.} (\bibinfo {collaboration} {Planck}),\ }\bibfield
  {title} {\enquote {\bibinfo {title} {{Planck 2018 results. VI. Cosmological
  parameters}},}\ }\href {\doibase 10.1051/0004-6361/201833910} {\bibfield
  {journal} {\bibinfo  {journal} {Astron. Astrophys.}\ }\textbf {\bibinfo
  {volume} {641}},\ \bibinfo {pages} {A6} (\bibinfo {year} {2020})},\ \bibinfo
  {note} {[Erratum: Astron.Astrophys. 652, C4 (2021)]},\ \Eprint
  {http://arxiv.org/abs/1807.06209} {arXiv:1807.06209 [astro-ph.CO]}
  \BibitemShut {NoStop}%
\bibitem [{\citenamefont {Baer}\ \emph {et~al.}(2015)\citenamefont {Baer},
  \citenamefont {Choi}, \citenamefont {Kim},\ and\ \citenamefont
  {Roszkowski}}]{Baer:2014eja}%
  \BibitemOpen
  \bibfield  {author} {\bibinfo {author} {\bibfnamefont {Howard}\ \bibnamefont
  {Baer}}, \bibinfo {author} {\bibfnamefont {Ki-Young}\ \bibnamefont {Choi}},
  \bibinfo {author} {\bibfnamefont {Jihn~E.}\ \bibnamefont {Kim}}, \ and\
  \bibinfo {author} {\bibfnamefont {Leszek}\ \bibnamefont {Roszkowski}},\
  }\bibfield  {title} {\enquote {\bibinfo {title} {{Dark matter production in
  the early Universe: beyond the thermal WIMP paradigm}},}\ }\href {\doibase
  10.1016/j.physrep.2014.10.002} {\bibfield  {journal} {\bibinfo  {journal}
  {Phys. Rept.}\ }\textbf {\bibinfo {volume} {555}},\ \bibinfo {pages} {1--60}
  (\bibinfo {year} {2015})},\ \Eprint {http://arxiv.org/abs/1407.0017}
  {arXiv:1407.0017 [hep-ph]} \BibitemShut {NoStop}%
\bibitem [{\citenamefont {Arcadi}\ \emph {et~al.}(2018)\citenamefont {Arcadi},
  \citenamefont {Dutra}, \citenamefont {Ghosh}, \citenamefont {Lindner},
  \citenamefont {Mambrini}, \citenamefont {Pierre}, \citenamefont {Profumo},\
  and\ \citenamefont {Queiroz}}]{Arcadi:2017kky}%
  \BibitemOpen
  \bibfield  {author} {\bibinfo {author} {\bibfnamefont {Giorgio}\ \bibnamefont
  {Arcadi}}, \bibinfo {author} {\bibfnamefont {Ma\'\i{}ra}\ \bibnamefont
  {Dutra}}, \bibinfo {author} {\bibfnamefont {Pradipta}\ \bibnamefont {Ghosh}},
  \bibinfo {author} {\bibfnamefont {Manfred}\ \bibnamefont {Lindner}}, \bibinfo
  {author} {\bibfnamefont {Yann}\ \bibnamefont {Mambrini}}, \bibinfo {author}
  {\bibfnamefont {Mathias}\ \bibnamefont {Pierre}}, \bibinfo {author}
  {\bibfnamefont {Stefano}\ \bibnamefont {Profumo}}, \ and\ \bibinfo {author}
  {\bibfnamefont {Farinaldo~S.}\ \bibnamefont {Queiroz}},\ }\bibfield  {title}
  {\enquote {\bibinfo {title} {{The waning of the WIMP? A review of models,
  searches, and constraints}},}\ }\href {\doibase
  10.1140/epjc/s10052-018-5662-y} {\bibfield  {journal} {\bibinfo  {journal}
  {Eur. Phys. J. C}\ }\textbf {\bibinfo {volume} {78}},\ \bibinfo {pages} {203}
  (\bibinfo {year} {2018})},\ \Eprint {http://arxiv.org/abs/1703.07364}
  {arXiv:1703.07364 [hep-ph]} \BibitemShut {NoStop}%
\bibitem [{\citenamefont {Roszkowski}\ \emph {et~al.}(2018)\citenamefont
  {Roszkowski}, \citenamefont {Sessolo},\ and\ \citenamefont
  {Trojanowski}}]{Roszkowski:2017nbc}%
  \BibitemOpen
  \bibfield  {author} {\bibinfo {author} {\bibfnamefont {Leszek}\ \bibnamefont
  {Roszkowski}}, \bibinfo {author} {\bibfnamefont {Enrico~Maria}\ \bibnamefont
  {Sessolo}}, \ and\ \bibinfo {author} {\bibfnamefont {Sebastian}\ \bibnamefont
  {Trojanowski}},\ }\bibfield  {title} {\enquote {\bibinfo {title} {{WIMP dark
  matter candidates and searches\textemdash{}current status and future
  prospects}},}\ }\href {\doibase 10.1088/1361-6633/aab913} {\bibfield
  {journal} {\bibinfo  {journal} {Rept. Prog. Phys.}\ }\textbf {\bibinfo
  {volume} {81}},\ \bibinfo {pages} {066201} (\bibinfo {year} {2018})},\
  \Eprint {http://arxiv.org/abs/1707.06277} {arXiv:1707.06277 [hep-ph]}
  \BibitemShut {NoStop}%
\bibitem [{\citenamefont {Han}\ \emph {et~al.}(2022)\citenamefont {Han},
  \citenamefont {Liu}, \citenamefont {Wang},\ and\ \citenamefont
  {Wang}}]{Han:2022ubw}%
  \BibitemOpen
  \bibfield  {author} {\bibinfo {author} {\bibfnamefont {Tao}\ \bibnamefont
  {Han}}, \bibinfo {author} {\bibfnamefont {Zhen}\ \bibnamefont {Liu}},
  \bibinfo {author} {\bibfnamefont {Lian-Tao}\ \bibnamefont {Wang}}, \ and\
  \bibinfo {author} {\bibfnamefont {Xing}\ \bibnamefont {Wang}},\ }\bibfield
  {title} {\enquote {\bibinfo {title} {{WIMP Dark Matter at High Energy Muon
  Colliders $-$A White Paper for Snowmass 2021}},}\ }in\ \href@noop {} {\emph
  {\bibinfo {booktitle} {{2022 Snowmass Summer Study}}}}\ (\bibinfo {year}
  {2022})\ \Eprint {http://arxiv.org/abs/2203.07351} {arXiv:2203.07351
  [hep-ph]} \BibitemShut {NoStop}%
\bibitem [{\citenamefont {Hall}\ \emph {et~al.}(2010)\citenamefont {Hall},
  \citenamefont {Jedamzik}, \citenamefont {March-Russell},\ and\ \citenamefont
  {West}}]{Hall:2009bx}%
  \BibitemOpen
  \bibfield  {author} {\bibinfo {author} {\bibfnamefont {Lawrence~J.}\
  \bibnamefont {Hall}}, \bibinfo {author} {\bibfnamefont {Karsten}\
  \bibnamefont {Jedamzik}}, \bibinfo {author} {\bibfnamefont {John}\
  \bibnamefont {March-Russell}}, \ and\ \bibinfo {author} {\bibfnamefont
  {Stephen~M.}\ \bibnamefont {West}},\ }\bibfield  {title} {\enquote {\bibinfo
  {title} {{Freeze-In Production of FIMP Dark Matter}},}\ }\href {\doibase
  10.1007/JHEP03(2010)080} {\bibfield  {journal} {\bibinfo  {journal} {JHEP}\
  }\textbf {\bibinfo {volume} {03}},\ \bibinfo {pages} {080} (\bibinfo {year}
  {2010})},\ \Eprint {http://arxiv.org/abs/0911.1120} {arXiv:0911.1120
  [hep-ph]} \BibitemShut {NoStop}%
\bibitem [{\citenamefont {Bernal}\ \emph {et~al.}(2017)\citenamefont {Bernal},
  \citenamefont {Heikinheimo}, \citenamefont {Tenkanen}, \citenamefont
  {Tuominen},\ and\ \citenamefont {Vaskonen}}]{Bernal:2017kxu}%
  \BibitemOpen
  \bibfield  {author} {\bibinfo {author} {\bibfnamefont {Nicol\'as}\
  \bibnamefont {Bernal}}, \bibinfo {author} {\bibfnamefont {Matti}\
  \bibnamefont {Heikinheimo}}, \bibinfo {author} {\bibfnamefont {Tommi}\
  \bibnamefont {Tenkanen}}, \bibinfo {author} {\bibfnamefont {Kimmo}\
  \bibnamefont {Tuominen}}, \ and\ \bibinfo {author} {\bibfnamefont {Ville}\
  \bibnamefont {Vaskonen}},\ }\bibfield  {title} {\enquote {\bibinfo {title}
  {{The Dawn of FIMP Dark Matter: A Review of Models and Constraints}},}\
  }\href {\doibase 10.1142/S0217751X1730023X} {\bibfield  {journal} {\bibinfo
  {journal} {Int. J. Mod. Phys. A}\ }\textbf {\bibinfo {volume} {32}},\
  \bibinfo {pages} {1730023} (\bibinfo {year} {2017})},\ \Eprint
  {http://arxiv.org/abs/1706.07442} {arXiv:1706.07442 [hep-ph]} \BibitemShut
  {NoStop}%
\bibitem [{\citenamefont {Chang}\ \emph {et~al.}(2021)\citenamefont {Chang},
  \citenamefont {Essig},\ and\ \citenamefont {Reinert}}]{Chang:2019xva}%
  \BibitemOpen
  \bibfield  {author} {\bibinfo {author} {\bibfnamefont {Jae~Hyeok}\
  \bibnamefont {Chang}}, \bibinfo {author} {\bibfnamefont {Rouven}\
  \bibnamefont {Essig}}, \ and\ \bibinfo {author} {\bibfnamefont {Annika}\
  \bibnamefont {Reinert}},\ }\bibfield  {title} {\enquote {\bibinfo {title}
  {{Light(ly)-coupled Dark Matter in the keV Range: Freeze-In and
  Constraints}},}\ }\href {\doibase 10.1007/JHEP03(2021)141} {\bibfield
  {journal} {\bibinfo  {journal} {JHEP}\ }\textbf {\bibinfo {volume} {03}},\
  \bibinfo {pages} {141} (\bibinfo {year} {2021})},\ \Eprint
  {http://arxiv.org/abs/1911.03389} {arXiv:1911.03389 [hep-ph]} \BibitemShut
  {NoStop}%
\bibitem [{\citenamefont {Dvorkin}\ \emph {et~al.}(2021)\citenamefont
  {Dvorkin}, \citenamefont {Lin},\ and\ \citenamefont
  {Schutz}}]{Dvorkin:2020xga}%
  \BibitemOpen
  \bibfield  {author} {\bibinfo {author} {\bibfnamefont {Cora}\ \bibnamefont
  {Dvorkin}}, \bibinfo {author} {\bibfnamefont {Tongyan}\ \bibnamefont {Lin}},
  \ and\ \bibinfo {author} {\bibfnamefont {Katelin}\ \bibnamefont {Schutz}},\
  }\bibfield  {title} {\enquote {\bibinfo {title} {{Cosmology of Sub-MeV Dark
  Matter Freeze-In}},}\ }\href {\doibase 10.1103/PhysRevLett.127.111301}
  {\bibfield  {journal} {\bibinfo  {journal} {Phys. Rev. Lett.}\ }\textbf
  {\bibinfo {volume} {127}},\ \bibinfo {pages} {111301} (\bibinfo {year}
  {2021})},\ \Eprint {http://arxiv.org/abs/2011.08186} {arXiv:2011.08186
  [astro-ph.CO]} \BibitemShut {NoStop}%
\bibitem [{\citenamefont {Coskuner}\ \emph {et~al.}(2022)\citenamefont
  {Coskuner}, \citenamefont {Trickle}, \citenamefont {Zhang},\ and\
  \citenamefont {Zurek}}]{Coskuner:2021qxo}%
  \BibitemOpen
  \bibfield  {author} {\bibinfo {author} {\bibfnamefont {Ahmet}\ \bibnamefont
  {Coskuner}}, \bibinfo {author} {\bibfnamefont {Tanner}\ \bibnamefont
  {Trickle}}, \bibinfo {author} {\bibfnamefont {Zhengkang}\ \bibnamefont
  {Zhang}}, \ and\ \bibinfo {author} {\bibfnamefont {Kathryn~M.}\ \bibnamefont
  {Zurek}},\ }\bibfield  {title} {\enquote {\bibinfo {title} {{Directional
  detectability of dark matter with single phonon excitations: Target
  comparison}},}\ }\href {\doibase 10.1103/PhysRevD.105.015010} {\bibfield
  {journal} {\bibinfo  {journal} {Phys. Rev. D}\ }\textbf {\bibinfo {volume}
  {105}},\ \bibinfo {pages} {015010} (\bibinfo {year} {2022})},\ \Eprint
  {http://arxiv.org/abs/2102.09567} {arXiv:2102.09567 [hep-ph]} \BibitemShut
  {NoStop}%
\bibitem [{\citenamefont {Chung}\ \emph {et~al.}(1998)\citenamefont {Chung},
  \citenamefont {Kolb},\ and\ \citenamefont {Riotto}}]{Chung:1998ua}%
  \BibitemOpen
  \bibfield  {author} {\bibinfo {author} {\bibfnamefont {Daniel J.~H.}\
  \bibnamefont {Chung}}, \bibinfo {author} {\bibfnamefont {Edward~W.}\
  \bibnamefont {Kolb}}, \ and\ \bibinfo {author} {\bibfnamefont {Antonio}\
  \bibnamefont {Riotto}},\ }\bibfield  {title} {\enquote {\bibinfo {title}
  {{Nonthermal supermassive dark matter}},}\ }\href {\doibase
  10.1103/PhysRevLett.81.4048} {\bibfield  {journal} {\bibinfo  {journal}
  {Phys. Rev. Lett.}\ }\textbf {\bibinfo {volume} {81}},\ \bibinfo {pages}
  {4048--4051} (\bibinfo {year} {1998})},\ \Eprint
  {http://arxiv.org/abs/hep-ph/9805473} {arXiv:hep-ph/9805473} \BibitemShut
  {NoStop}%
\bibitem [{\citenamefont {Carney}\ \emph {et~al.}(2022)\citenamefont {Carney}
  \emph {et~al.}}]{Carney:2022gse}%
  \BibitemOpen
  \bibfield  {author} {\bibinfo {author} {\bibfnamefont {Daniel}\ \bibnamefont
  {Carney}} \emph {et~al.},\ }\bibfield  {title} {\enquote {\bibinfo {title}
  {{Snowmass2021 Cosmic Frontier White Paper: Ultraheavy particle dark
  matter}},}\ \ }(\bibinfo {year} {2022})\ \Eprint
  {http://arxiv.org/abs/2203.06508} {arXiv:2203.06508 [hep-ph]} \BibitemShut
  {NoStop}%
\bibitem [{\citenamefont {Essig}\ \emph {et~al.}(2012)\citenamefont {Essig},
  \citenamefont {Mardon},\ and\ \citenamefont {Volansky}}]{Essig:2011nj}%
  \BibitemOpen
  \bibfield  {author} {\bibinfo {author} {\bibfnamefont {Rouven}\ \bibnamefont
  {Essig}}, \bibinfo {author} {\bibfnamefont {Jeremy}\ \bibnamefont {Mardon}},
  \ and\ \bibinfo {author} {\bibfnamefont {Tomer}\ \bibnamefont {Volansky}},\
  }\bibfield  {title} {\enquote {\bibinfo {title} {{Direct Detection of Sub-GeV
  Dark Matter}},}\ }\href {\doibase 10.1103/PhysRevD.85.076007} {\bibfield
  {journal} {\bibinfo  {journal} {Phys. Rev. D}\ }\textbf {\bibinfo {volume}
  {85}},\ \bibinfo {pages} {076007} (\bibinfo {year} {2012})},\ \Eprint
  {http://arxiv.org/abs/1108.5383} {arXiv:1108.5383 [hep-ph]} \BibitemShut
  {NoStop}%
\bibitem [{\citenamefont {Chu}\ \emph {et~al.}(2012)\citenamefont {Chu},
  \citenamefont {Hambye},\ and\ \citenamefont {Tytgat}}]{Chu:2011be}%
  \BibitemOpen
  \bibfield  {author} {\bibinfo {author} {\bibfnamefont {Xiaoyong}\
  \bibnamefont {Chu}}, \bibinfo {author} {\bibfnamefont {Thomas}\ \bibnamefont
  {Hambye}}, \ and\ \bibinfo {author} {\bibfnamefont {Michel H.~G.}\
  \bibnamefont {Tytgat}},\ }\bibfield  {title} {\enquote {\bibinfo {title}
  {{The Four Basic Ways of Creating Dark Matter Through a Portal}},}\ }\href
  {\doibase 10.1088/1475-7516/2012/05/034} {\bibfield  {journal} {\bibinfo
  {journal} {JCAP}\ }\textbf {\bibinfo {volume} {05}},\ \bibinfo {pages} {034}
  (\bibinfo {year} {2012})},\ \Eprint {http://arxiv.org/abs/1112.0493}
  {arXiv:1112.0493 [hep-ph]} \BibitemShut {NoStop}%
\bibitem [{\citenamefont {Essig}\ \emph {et~al.}(2016)\citenamefont {Essig},
  \citenamefont {Fernandez-Serra}, \citenamefont {Mardon}, \citenamefont
  {Soto}, \citenamefont {Volansky},\ and\ \citenamefont {Yu}}]{Essig:2015cda}%
  \BibitemOpen
  \bibfield  {author} {\bibinfo {author} {\bibfnamefont {Rouven}\ \bibnamefont
  {Essig}}, \bibinfo {author} {\bibfnamefont {Marivi}\ \bibnamefont
  {Fernandez-Serra}}, \bibinfo {author} {\bibfnamefont {Jeremy}\ \bibnamefont
  {Mardon}}, \bibinfo {author} {\bibfnamefont {Adrian}\ \bibnamefont {Soto}},
  \bibinfo {author} {\bibfnamefont {Tomer}\ \bibnamefont {Volansky}}, \ and\
  \bibinfo {author} {\bibfnamefont {Tien-Tien}\ \bibnamefont {Yu}},\ }\bibfield
   {title} {\enquote {\bibinfo {title} {{Direct Detection of sub-GeV Dark
  Matter with Semiconductor Targets}},}\ }\href {\doibase
  10.1007/JHEP05(2016)046} {\bibfield  {journal} {\bibinfo  {journal} {JHEP}\
  }\textbf {\bibinfo {volume} {05}},\ \bibinfo {pages} {046} (\bibinfo {year}
  {2016})},\ \Eprint {http://arxiv.org/abs/1509.01598} {arXiv:1509.01598
  [hep-ph]} \BibitemShut {NoStop}%
\bibitem [{\citenamefont {Hambye}\ \emph {et~al.}(2018)\citenamefont {Hambye},
  \citenamefont {Tytgat}, \citenamefont {Vandecasteele},\ and\ \citenamefont
  {Vanderheyden}}]{Hambye:2018dpi}%
  \BibitemOpen
  \bibfield  {author} {\bibinfo {author} {\bibfnamefont {Thomas}\ \bibnamefont
  {Hambye}}, \bibinfo {author} {\bibfnamefont {Michel H.~G.}\ \bibnamefont
  {Tytgat}}, \bibinfo {author} {\bibfnamefont {J\'er\^ome}\ \bibnamefont
  {Vandecasteele}}, \ and\ \bibinfo {author} {\bibfnamefont {Laurent}\
  \bibnamefont {Vanderheyden}},\ }\bibfield  {title} {\enquote {\bibinfo
  {title} {{Dark matter direct detection is testing freeze-in}},}\ }\href
  {\doibase 10.1103/PhysRevD.98.075017} {\bibfield  {journal} {\bibinfo
  {journal} {Phys. Rev. D}\ }\textbf {\bibinfo {volume} {98}},\ \bibinfo
  {pages} {075017} (\bibinfo {year} {2018})},\ \Eprint
  {http://arxiv.org/abs/1807.05022} {arXiv:1807.05022 [hep-ph]} \BibitemShut
  {NoStop}%
\bibitem [{\citenamefont {Heeba}\ and\ \citenamefont
  {Kahlhoefer}(2020)}]{Heeba:2019jho}%
  \BibitemOpen
  \bibfield  {author} {\bibinfo {author} {\bibfnamefont {Saniya}\ \bibnamefont
  {Heeba}}\ and\ \bibinfo {author} {\bibfnamefont {Felix}\ \bibnamefont
  {Kahlhoefer}},\ }\bibfield  {title} {\enquote {\bibinfo {title} {{Probing the
  freeze-in mechanism in dark matter models with U(1)' gauge extensions}},}\
  }\href {\doibase 10.1103/PhysRevD.101.035043} {\bibfield  {journal} {\bibinfo
   {journal} {Phys. Rev. D}\ }\textbf {\bibinfo {volume} {101}},\ \bibinfo
  {pages} {035043} (\bibinfo {year} {2020})},\ \Eprint
  {http://arxiv.org/abs/1908.09834} {arXiv:1908.09834 [hep-ph]} \BibitemShut
  {NoStop}%
\bibitem [{\citenamefont {An}\ and\ \citenamefont {Yang}(2021)}]{An:2020tcg}%
  \BibitemOpen
  \bibfield  {author} {\bibinfo {author} {\bibfnamefont {Haipeng}\ \bibnamefont
  {An}}\ and\ \bibinfo {author} {\bibfnamefont {Daneng}\ \bibnamefont {Yang}},\
  }\bibfield  {title} {\enquote {\bibinfo {title} {{Direct detection of
  freeze-in inelastic dark matter}},}\ }\href {\doibase
  10.1016/j.physletb.2021.136408} {\bibfield  {journal} {\bibinfo  {journal}
  {Phys. Lett. B}\ }\textbf {\bibinfo {volume} {818}},\ \bibinfo {pages}
  {136408} (\bibinfo {year} {2021})},\ \Eprint
  {http://arxiv.org/abs/2006.15672} {arXiv:2006.15672 [hep-ph]} \BibitemShut
  {NoStop}%
\bibitem [{\citenamefont {Cosme}\ \emph
  {et~al.}(2021{\natexlab{a}})\citenamefont {Cosme}, \citenamefont {Dutra},
  \citenamefont {Godfrey},\ and\ \citenamefont {Gray}}]{Cosme:2021baj}%
  \BibitemOpen
  \bibfield  {author} {\bibinfo {author} {\bibfnamefont {Catarina}\
  \bibnamefont {Cosme}}, \bibinfo {author} {\bibfnamefont {Ma\'\i{}ra}\
  \bibnamefont {Dutra}}, \bibinfo {author} {\bibfnamefont {Stephen}\
  \bibnamefont {Godfrey}}, \ and\ \bibinfo {author} {\bibfnamefont {Taylor~R.}\
  \bibnamefont {Gray}},\ }\bibfield  {title} {\enquote {\bibinfo {title}
  {{Testing freeze-in with axial and vector $Z'$ bosons}},}\ }\href {\doibase
  10.1007/JHEP09(2021)056} {\bibfield  {journal} {\bibinfo  {journal} {JHEP}\
  }\textbf {\bibinfo {volume} {09}},\ \bibinfo {pages} {056} (\bibinfo {year}
  {2021}{\natexlab{a}})},\ \Eprint {http://arxiv.org/abs/2104.13937}
  {arXiv:2104.13937 [hep-ph]} \BibitemShut {NoStop}%
\bibitem [{\citenamefont {Brdar}\ \emph {et~al.}(2018)\citenamefont {Brdar},
  \citenamefont {Kopp}, \citenamefont {Liu},\ and\ \citenamefont
  {Wang}}]{Brdar:2017wgy}%
  \BibitemOpen
  \bibfield  {author} {\bibinfo {author} {\bibfnamefont {Vedran}\ \bibnamefont
  {Brdar}}, \bibinfo {author} {\bibfnamefont {Joachim}\ \bibnamefont {Kopp}},
  \bibinfo {author} {\bibfnamefont {Jia}\ \bibnamefont {Liu}}, \ and\ \bibinfo
  {author} {\bibfnamefont {Xiao-Ping}\ \bibnamefont {Wang}},\ }\bibfield
  {title} {\enquote {\bibinfo {title} {{X-Ray Lines from Dark Matter
  Annihilation at the keV Scale}},}\ }\href {\doibase
  10.1103/PhysRevLett.120.061301} {\bibfield  {journal} {\bibinfo  {journal}
  {Phys. Rev. Lett.}\ }\textbf {\bibinfo {volume} {120}},\ \bibinfo {pages}
  {061301} (\bibinfo {year} {2018})},\ \Eprint
  {http://arxiv.org/abs/1710.02146} {arXiv:1710.02146 [hep-ph]} \BibitemShut
  {NoStop}%
\bibitem [{\citenamefont {Biswas}\ \emph {et~al.}(2020)\citenamefont {Biswas},
  \citenamefont {Ganguly},\ and\ \citenamefont {Roy}}]{Biswas:2019iqm}%
  \BibitemOpen
  \bibfield  {author} {\bibinfo {author} {\bibfnamefont {Anirban}\ \bibnamefont
  {Biswas}}, \bibinfo {author} {\bibfnamefont {Sougata}\ \bibnamefont
  {Ganguly}}, \ and\ \bibinfo {author} {\bibfnamefont {Sourov}\ \bibnamefont
  {Roy}},\ }\bibfield  {title} {\enquote {\bibinfo {title} {{Fermionic dark
  matter via UV and IR freeze-in and its possible X-ray signature}},}\ }\href
  {\doibase 10.1088/1475-7516/2020/03/043} {\bibfield  {journal} {\bibinfo
  {journal} {JCAP}\ }\textbf {\bibinfo {volume} {03}},\ \bibinfo {pages} {043}
  (\bibinfo {year} {2020})},\ \Eprint {http://arxiv.org/abs/1907.07973}
  {arXiv:1907.07973 [hep-ph]} \BibitemShut {NoStop}%
\bibitem [{\citenamefont {Bernal}\ \emph {et~al.}(2019)\citenamefont {Bernal},
  \citenamefont {Cosme},\ and\ \citenamefont {Tenkanen}}]{Bernal:2018ins}%
  \BibitemOpen
  \bibfield  {author} {\bibinfo {author} {\bibfnamefont {Nicol\'as}\
  \bibnamefont {Bernal}}, \bibinfo {author} {\bibfnamefont {Catarina}\
  \bibnamefont {Cosme}}, \ and\ \bibinfo {author} {\bibfnamefont {Tommi}\
  \bibnamefont {Tenkanen}},\ }\bibfield  {title} {\enquote {\bibinfo {title}
  {{Phenomenology of Self-Interacting Dark Matter in a Matter-Dominated
  Universe}},}\ }\href {\doibase 10.1140/epjc/s10052-019-6608-8} {\bibfield
  {journal} {\bibinfo  {journal} {Eur. Phys. J. C}\ }\textbf {\bibinfo {volume}
  {79}},\ \bibinfo {pages} {99} (\bibinfo {year} {2019})},\ \Eprint
  {http://arxiv.org/abs/1803.08064} {arXiv:1803.08064 [hep-ph]} \BibitemShut
  {NoStop}%
\bibitem [{\citenamefont {Cosme}\ \emph
  {et~al.}(2021{\natexlab{b}})\citenamefont {Cosme}, \citenamefont {Dutra},
  \citenamefont {Ma}, \citenamefont {Wu},\ and\ \citenamefont
  {Yang}}]{Cosme:2020mck}%
  \BibitemOpen
  \bibfield  {author} {\bibinfo {author} {\bibfnamefont {Catarina}\
  \bibnamefont {Cosme}}, \bibinfo {author} {\bibfnamefont {Ma\'\i{}ra}\
  \bibnamefont {Dutra}}, \bibinfo {author} {\bibfnamefont {Teng}\ \bibnamefont
  {Ma}}, \bibinfo {author} {\bibfnamefont {Yongcheng}\ \bibnamefont {Wu}}, \
  and\ \bibinfo {author} {\bibfnamefont {Litao}\ \bibnamefont {Yang}},\
  }\bibfield  {title} {\enquote {\bibinfo {title} {{Neutrino portal to FIMP
  dark matter with an early matter era}},}\ }\href {\doibase
  10.1007/JHEP03(2021)026} {\bibfield  {journal} {\bibinfo  {journal} {JHEP}\
  }\textbf {\bibinfo {volume} {03}},\ \bibinfo {pages} {026} (\bibinfo {year}
  {2021}{\natexlab{b}})},\ \Eprint {http://arxiv.org/abs/2003.01723}
  {arXiv:2003.01723 [hep-ph]} \BibitemShut {NoStop}%
\bibitem [{\citenamefont {Kahlhoefer}(2017)}]{Kahlhoefer:2017dnp}%
  \BibitemOpen
  \bibfield  {author} {\bibinfo {author} {\bibfnamefont {Felix}\ \bibnamefont
  {Kahlhoefer}},\ }\bibfield  {title} {\enquote {\bibinfo {title} {{Review of
  LHC Dark Matter Searches}},}\ }\href {\doibase 10.1142/S0217751X1730006X}
  {\bibfield  {journal} {\bibinfo  {journal} {Int. J. Mod. Phys. A}\ }\textbf
  {\bibinfo {volume} {32}},\ \bibinfo {pages} {1730006} (\bibinfo {year}
  {2017})},\ \Eprint {http://arxiv.org/abs/1702.02430} {arXiv:1702.02430
  [hep-ph]} \BibitemShut {NoStop}%
\bibitem [{\citenamefont {Trevisani}(2018)}]{Trevisani:2018psx}%
  \BibitemOpen
  \bibfield  {author} {\bibinfo {author} {\bibfnamefont {Nicol\`o}\
  \bibnamefont {Trevisani}} (\bibinfo {collaboration} {ATLAS, CMS}),\
  }\bibfield  {title} {\enquote {\bibinfo {title} {{Collider Searches for Dark
  Matter (ATLAS + CMS)}},}\ }\href {\doibase 10.3390/universe4110131}
  {\bibfield  {journal} {\bibinfo  {journal} {Universe}\ }\textbf {\bibinfo
  {volume} {4}},\ \bibinfo {pages} {131} (\bibinfo {year} {2018})}\BibitemShut
  {NoStop}%
\bibitem [{\citenamefont {Diehl}\ \emph {et~al.}(1995)\citenamefont {Diehl},
  \citenamefont {Kane}, \citenamefont {Kolda},\ and\ \citenamefont
  {Wells}}]{Diehl:1994ff}%
  \BibitemOpen
  \bibfield  {author} {\bibinfo {author} {\bibfnamefont {E.}~\bibnamefont
  {Diehl}}, \bibinfo {author} {\bibfnamefont {Gordon~L.}\ \bibnamefont {Kane}},
  \bibinfo {author} {\bibfnamefont {Christopher~F.}\ \bibnamefont {Kolda}}, \
  and\ \bibinfo {author} {\bibfnamefont {James~D.}\ \bibnamefont {Wells}},\
  }\bibfield  {title} {\enquote {\bibinfo {title} {{Theory, phenomenology, and
  prospects for detection of supersymmetric dark matter}},}\ }\href {\doibase
  10.1103/PhysRevD.52.4223} {\bibfield  {journal} {\bibinfo  {journal} {Phys.
  Rev. D}\ }\textbf {\bibinfo {volume} {52}},\ \bibinfo {pages} {4223--4239}
  (\bibinfo {year} {1995})},\ \Eprint {http://arxiv.org/abs/hep-ph/9502399}
  {arXiv:hep-ph/9502399} \BibitemShut {NoStop}%
\bibitem [{\citenamefont {Goodman}\ \emph {et~al.}(2010)\citenamefont
  {Goodman}, \citenamefont {Ibe}, \citenamefont {Rajaraman}, \citenamefont
  {Shepherd}, \citenamefont {Tait},\ and\ \citenamefont {Yu}}]{Goodman:2010ku}%
  \BibitemOpen
  \bibfield  {author} {\bibinfo {author} {\bibfnamefont {Jessica}\ \bibnamefont
  {Goodman}}, \bibinfo {author} {\bibfnamefont {Masahiro}\ \bibnamefont {Ibe}},
  \bibinfo {author} {\bibfnamefont {Arvind}\ \bibnamefont {Rajaraman}},
  \bibinfo {author} {\bibfnamefont {William}\ \bibnamefont {Shepherd}},
  \bibinfo {author} {\bibfnamefont {Tim M.~P.}\ \bibnamefont {Tait}}, \ and\
  \bibinfo {author} {\bibfnamefont {Hai-Bo}\ \bibnamefont {Yu}},\ }\bibfield
  {title} {\enquote {\bibinfo {title} {{Constraints on Dark Matter from
  Colliders}},}\ }\href {\doibase 10.1103/PhysRevD.82.116010} {\bibfield
  {journal} {\bibinfo  {journal} {Phys. Rev. D}\ }\textbf {\bibinfo {volume}
  {82}},\ \bibinfo {pages} {116010} (\bibinfo {year} {2010})},\ \Eprint
  {http://arxiv.org/abs/1008.1783} {arXiv:1008.1783 [hep-ph]} \BibitemShut
  {NoStop}%
\bibitem [{\citenamefont {Fox}\ \emph {et~al.}(2012)\citenamefont {Fox},
  \citenamefont {Harnik}, \citenamefont {Kopp},\ and\ \citenamefont
  {Tsai}}]{Fox:2011pm}%
  \BibitemOpen
  \bibfield  {author} {\bibinfo {author} {\bibfnamefont {Patrick~J.}\
  \bibnamefont {Fox}}, \bibinfo {author} {\bibfnamefont {Roni}\ \bibnamefont
  {Harnik}}, \bibinfo {author} {\bibfnamefont {Joachim}\ \bibnamefont {Kopp}},
  \ and\ \bibinfo {author} {\bibfnamefont {Yuhsin}\ \bibnamefont {Tsai}},\
  }\bibfield  {title} {\enquote {\bibinfo {title} {{Missing Energy Signatures
  of Dark Matter at the LHC}},}\ }\href {\doibase 10.1103/PhysRevD.85.056011}
  {\bibfield  {journal} {\bibinfo  {journal} {Phys. Rev. D}\ }\textbf {\bibinfo
  {volume} {85}},\ \bibinfo {pages} {056011} (\bibinfo {year} {2012})},\
  \Eprint {http://arxiv.org/abs/1109.4398} {arXiv:1109.4398 [hep-ph]}
  \BibitemShut {NoStop}%
\bibitem [{\citenamefont {Aad}\ \emph {et~al.}(2021)\citenamefont {Aad} \emph
  {et~al.}}]{ATLAS:2021kxv}%
  \BibitemOpen
  \bibfield  {author} {\bibinfo {author} {\bibfnamefont {Georges}\ \bibnamefont
  {Aad}} \emph {et~al.} (\bibinfo {collaboration} {ATLAS}),\ }\bibfield
  {title} {\enquote {\bibinfo {title} {{Search for new phenomena in events with
  an energetic jet and missing transverse momentum in $pp$ collisions at $\sqrt
  {s}$ =13 TeV with the ATLAS detector}},}\ }\href {\doibase
  10.1103/PhysRevD.103.112006} {\bibfield  {journal} {\bibinfo  {journal}
  {Phys. Rev. D}\ }\textbf {\bibinfo {volume} {103}},\ \bibinfo {pages}
  {112006} (\bibinfo {year} {2021})},\ \Eprint
  {http://arxiv.org/abs/2102.10874} {arXiv:2102.10874 [hep-ex]} \BibitemShut
  {NoStop}%
\bibitem [{\citenamefont {Tumasyan}\ \emph {et~al.}(2021)\citenamefont
  {Tumasyan} \emph {et~al.}}]{CMS:2021far}%
  \BibitemOpen
  \bibfield  {author} {\bibinfo {author} {\bibfnamefont {Armen}\ \bibnamefont
  {Tumasyan}} \emph {et~al.} (\bibinfo {collaboration} {CMS}),\ }\bibfield
  {title} {\enquote {\bibinfo {title} {{Search for new particles in events with
  energetic jets and large missing transverse momentum in proton-proton
  collisions at $ \sqrt{s} $ = 13 TeV}},}\ }\href {\doibase
  10.1007/JHEP11(2021)153} {\bibfield  {journal} {\bibinfo  {journal} {JHEP}\
  }\textbf {\bibinfo {volume} {11}},\ \bibinfo {pages} {153} (\bibinfo {year}
  {2021})},\ \Eprint {http://arxiv.org/abs/2107.13021} {arXiv:2107.13021
  [hep-ex]} \BibitemShut {NoStop}%
\bibitem [{\citenamefont {Co}\ \emph {et~al.}(2015)\citenamefont {Co},
  \citenamefont {D'Eramo}, \citenamefont {Hall},\ and\ \citenamefont
  {Pappadopulo}}]{Co:2015pka}%
  \BibitemOpen
  \bibfield  {author} {\bibinfo {author} {\bibfnamefont {Raymond~T.}\
  \bibnamefont {Co}}, \bibinfo {author} {\bibfnamefont {Francesco}\
  \bibnamefont {D'Eramo}}, \bibinfo {author} {\bibfnamefont {Lawrence~J.}\
  \bibnamefont {Hall}}, \ and\ \bibinfo {author} {\bibfnamefont {Duccio}\
  \bibnamefont {Pappadopulo}},\ }\bibfield  {title} {\enquote {\bibinfo {title}
  {{Freeze-In Dark Matter with Displaced Signatures at Colliders}},}\ }\href
  {\doibase 10.1088/1475-7516/2015/12/024} {\bibfield  {journal} {\bibinfo
  {journal} {JCAP}\ }\textbf {\bibinfo {volume} {12}},\ \bibinfo {pages} {024}
  (\bibinfo {year} {2015})},\ \Eprint {http://arxiv.org/abs/1506.07532}
  {arXiv:1506.07532 [hep-ph]} \BibitemShut {NoStop}%
\bibitem [{\citenamefont {Hessler}\ \emph {et~al.}(2017)\citenamefont
  {Hessler}, \citenamefont {Ibarra}, \citenamefont {Molinaro},\ and\
  \citenamefont {Vogl}}]{Hessler:2016kwm}%
  \BibitemOpen
  \bibfield  {author} {\bibinfo {author} {\bibfnamefont {Andre~G.}\
  \bibnamefont {Hessler}}, \bibinfo {author} {\bibfnamefont {Alejandro}\
  \bibnamefont {Ibarra}}, \bibinfo {author} {\bibfnamefont {Emiliano}\
  \bibnamefont {Molinaro}}, \ and\ \bibinfo {author} {\bibfnamefont {Stefan}\
  \bibnamefont {Vogl}},\ }\bibfield  {title} {\enquote {\bibinfo {title}
  {{Probing the scotogenic FIMP at the LHC}},}\ }\href {\doibase
  10.1007/JHEP01(2017)100} {\bibfield  {journal} {\bibinfo  {journal} {JHEP}\
  }\textbf {\bibinfo {volume} {01}},\ \bibinfo {pages} {100} (\bibinfo {year}
  {2017})},\ \Eprint {http://arxiv.org/abs/1611.09540} {arXiv:1611.09540
  [hep-ph]} \BibitemShut {NoStop}%
\bibitem [{\citenamefont {Calibbi}\ \emph {et~al.}(2018)\citenamefont
  {Calibbi}, \citenamefont {Lopez-Honorez}, \citenamefont {Lowette},\ and\
  \citenamefont {Mariotti}}]{Calibbi:2018fqf}%
  \BibitemOpen
  \bibfield  {author} {\bibinfo {author} {\bibfnamefont {Lorenzo}\ \bibnamefont
  {Calibbi}}, \bibinfo {author} {\bibfnamefont {Laura}\ \bibnamefont
  {Lopez-Honorez}}, \bibinfo {author} {\bibfnamefont {Steven}\ \bibnamefont
  {Lowette}}, \ and\ \bibinfo {author} {\bibfnamefont {Alberto}\ \bibnamefont
  {Mariotti}},\ }\bibfield  {title} {\enquote {\bibinfo {title}
  {{Singlet-Doublet Dark Matter Freeze-in: LHC displaced signatures versus
  cosmology}},}\ }\href {\doibase 10.1007/JHEP09(2018)037} {\bibfield
  {journal} {\bibinfo  {journal} {JHEP}\ }\textbf {\bibinfo {volume} {09}},\
  \bibinfo {pages} {037} (\bibinfo {year} {2018})},\ \Eprint
  {http://arxiv.org/abs/1805.04423} {arXiv:1805.04423 [hep-ph]} \BibitemShut
  {NoStop}%
\bibitem [{\citenamefont {B\'elanger}\ \emph {et~al.}(2019)\citenamefont
  {B\'elanger} \emph {et~al.}}]{Belanger:2018sti}%
  \BibitemOpen
  \bibfield  {author} {\bibinfo {author} {\bibfnamefont {G.}~\bibnamefont
  {B\'elanger}} \emph {et~al.},\ }\bibfield  {title} {\enquote {\bibinfo
  {title} {{LHC-friendly minimal freeze-in models}},}\ }\href {\doibase
  10.1007/JHEP02(2019)186} {\bibfield  {journal} {\bibinfo  {journal} {JHEP}\
  }\textbf {\bibinfo {volume} {02}},\ \bibinfo {pages} {186} (\bibinfo {year}
  {2019})},\ \Eprint {http://arxiv.org/abs/1811.05478} {arXiv:1811.05478
  [hep-ph]} \BibitemShut {NoStop}%
\bibitem [{\citenamefont {No}\ \emph {et~al.}(2020)\citenamefont {No},
  \citenamefont {Tunney},\ and\ \citenamefont {Zaldivar}}]{No:2019gvl}%
  \BibitemOpen
  \bibfield  {author} {\bibinfo {author} {\bibfnamefont {Jose~Miguel}\
  \bibnamefont {No}}, \bibinfo {author} {\bibfnamefont {Patrick}\ \bibnamefont
  {Tunney}}, \ and\ \bibinfo {author} {\bibfnamefont {Bryan}\ \bibnamefont
  {Zaldivar}},\ }\bibfield  {title} {\enquote {\bibinfo {title} {{Probing Dark
  Matter freeze-in with long-lived particle signatures: MATHUSLA, HL-LHC and
  FCC-hh}},}\ }\href {\doibase 10.1007/JHEP03(2020)022} {\bibfield  {journal}
  {\bibinfo  {journal} {JHEP}\ }\textbf {\bibinfo {volume} {03}},\ \bibinfo
  {pages} {022} (\bibinfo {year} {2020})},\ \Eprint
  {http://arxiv.org/abs/1908.11387} {arXiv:1908.11387 [hep-ph]} \BibitemShut
  {NoStop}%
\bibitem [{\citenamefont {Okada}\ \emph {et~al.}(2020)\citenamefont {Okada},
  \citenamefont {Okada},\ and\ \citenamefont {Shafi}}]{Okada:2020cue}%
  \BibitemOpen
  \bibfield  {author} {\bibinfo {author} {\bibfnamefont {Nobuchika}\
  \bibnamefont {Okada}}, \bibinfo {author} {\bibfnamefont {Satomi}\
  \bibnamefont {Okada}}, \ and\ \bibinfo {author} {\bibfnamefont {Qaisar}\
  \bibnamefont {Shafi}},\ }\bibfield  {title} {\enquote {\bibinfo {title}
  {{Light $Z'$ and dark matter from U(1)$_X$ gauge symmetry}},}\ }\href
  {\doibase 10.1016/j.physletb.2020.135845} {\bibfield  {journal} {\bibinfo
  {journal} {Phys. Lett. B}\ }\textbf {\bibinfo {volume} {810}},\ \bibinfo
  {pages} {135845} (\bibinfo {year} {2020})},\ \Eprint
  {http://arxiv.org/abs/2003.02667} {arXiv:2003.02667 [hep-ph]} \BibitemShut
  {NoStop}%
\bibitem [{\citenamefont {Antoniadis}\ \emph {et~al.}(2010)\citenamefont
  {Antoniadis}, \citenamefont {Boyarsky}, \citenamefont {Espahbodi},
  \citenamefont {Ruchayskiy},\ and\ \citenamefont {Wells}}]{Antoniadis:2009ze}%
  \BibitemOpen
  \bibfield  {author} {\bibinfo {author} {\bibfnamefont {Ignatios}\
  \bibnamefont {Antoniadis}}, \bibinfo {author} {\bibfnamefont {Alexey}\
  \bibnamefont {Boyarsky}}, \bibinfo {author} {\bibfnamefont {Sam}\
  \bibnamefont {Espahbodi}}, \bibinfo {author} {\bibfnamefont {Oleg}\
  \bibnamefont {Ruchayskiy}}, \ and\ \bibinfo {author} {\bibfnamefont
  {James~D.}\ \bibnamefont {Wells}},\ }\bibfield  {title} {\enquote {\bibinfo
  {title} {{Anomaly driven signatures of new invisible physics at the Large
  Hadron Collider}},}\ }\href {\doibase 10.1016/j.nuclphysb.2009.09.009}
  {\bibfield  {journal} {\bibinfo  {journal} {Nucl. Phys. B}\ }\textbf
  {\bibinfo {volume} {824}},\ \bibinfo {pages} {296--313} (\bibinfo {year}
  {2010})},\ \Eprint {http://arxiv.org/abs/0901.0639} {arXiv:0901.0639
  [hep-ph]} \BibitemShut {NoStop}%
\bibitem [{\citenamefont {Dudas}\ \emph {et~al.}(2009)\citenamefont {Dudas},
  \citenamefont {Mambrini}, \citenamefont {Pokorski},\ and\ \citenamefont
  {Romagnoni}}]{Dudas:2009uq}%
  \BibitemOpen
  \bibfield  {author} {\bibinfo {author} {\bibfnamefont {E.}~\bibnamefont
  {Dudas}}, \bibinfo {author} {\bibfnamefont {Y.}~\bibnamefont {Mambrini}},
  \bibinfo {author} {\bibfnamefont {S.}~\bibnamefont {Pokorski}}, \ and\
  \bibinfo {author} {\bibfnamefont {A.}~\bibnamefont {Romagnoni}},\ }\bibfield
  {title} {\enquote {\bibinfo {title} {{(In)visible Z-prime and dark
  matter}},}\ }\href {\doibase 10.1088/1126-6708/2009/08/014} {\bibfield
  {journal} {\bibinfo  {journal} {JHEP}\ }\textbf {\bibinfo {volume} {08}},\
  \bibinfo {pages} {014} (\bibinfo {year} {2009})},\ \Eprint
  {http://arxiv.org/abs/0904.1745} {arXiv:0904.1745 [hep-ph]} \BibitemShut
  {NoStop}%
\bibitem [{\citenamefont {Dudas}\ \emph {et~al.}(2012)\citenamefont {Dudas},
  \citenamefont {Mambrini}, \citenamefont {Pokorski},\ and\ \citenamefont
  {Romagnoni}}]{Dudas:2012pb}%
  \BibitemOpen
  \bibfield  {author} {\bibinfo {author} {\bibfnamefont {Emilian}\ \bibnamefont
  {Dudas}}, \bibinfo {author} {\bibfnamefont {Yann}\ \bibnamefont {Mambrini}},
  \bibinfo {author} {\bibfnamefont {Stefan}\ \bibnamefont {Pokorski}}, \ and\
  \bibinfo {author} {\bibfnamefont {Alberto}\ \bibnamefont {Romagnoni}},\
  }\bibfield  {title} {\enquote {\bibinfo {title} {{Extra U(1) as natural
  source of a monochromatic gamma ray line}},}\ }\href {\doibase
  10.1007/JHEP10(2012)123} {\bibfield  {journal} {\bibinfo  {journal} {JHEP}\
  }\textbf {\bibinfo {volume} {10}},\ \bibinfo {pages} {123} (\bibinfo {year}
  {2012})},\ \Eprint {http://arxiv.org/abs/1205.1520} {arXiv:1205.1520
  [hep-ph]} \BibitemShut {NoStop}%
\bibitem [{\citenamefont {Arcadi}\ \emph {et~al.}(2017)\citenamefont {Arcadi},
  \citenamefont {Ghosh}, \citenamefont {Mambrini}, \citenamefont {Pierre},\
  and\ \citenamefont {Queiroz}}]{Arcadi:2017jqd}%
  \BibitemOpen
  \bibfield  {author} {\bibinfo {author} {\bibfnamefont {Giorgio}\ \bibnamefont
  {Arcadi}}, \bibinfo {author} {\bibfnamefont {Pradipta}\ \bibnamefont
  {Ghosh}}, \bibinfo {author} {\bibfnamefont {Yann}\ \bibnamefont {Mambrini}},
  \bibinfo {author} {\bibfnamefont {Mathias}\ \bibnamefont {Pierre}}, \ and\
  \bibinfo {author} {\bibfnamefont {Farinaldo~S.}\ \bibnamefont {Queiroz}},\
  }\bibfield  {title} {\enquote {\bibinfo {title} {{$Z'$ portal to Chern-Simons
  Dark Matter}},}\ }\href {\doibase 10.1088/1475-7516/2017/11/020} {\bibfield
  {journal} {\bibinfo  {journal} {JCAP}\ }\textbf {\bibinfo {volume} {11}},\
  \bibinfo {pages} {020} (\bibinfo {year} {2017})},\ \Eprint
  {http://arxiv.org/abs/1706.04198} {arXiv:1706.04198 [hep-ph]} \BibitemShut
  {NoStop}%
\bibitem [{\citenamefont {Dudas}\ \emph {et~al.}(2013)\citenamefont {Dudas},
  \citenamefont {Heurtier}, \citenamefont {Mambrini},\ and\ \citenamefont
  {Zaldivar}}]{Dudas:2013sia}%
  \BibitemOpen
  \bibfield  {author} {\bibinfo {author} {\bibfnamefont {Emilian}\ \bibnamefont
  {Dudas}}, \bibinfo {author} {\bibfnamefont {Lucien}\ \bibnamefont
  {Heurtier}}, \bibinfo {author} {\bibfnamefont {Yann}\ \bibnamefont
  {Mambrini}}, \ and\ \bibinfo {author} {\bibfnamefont {Bryan}\ \bibnamefont
  {Zaldivar}},\ }\bibfield  {title} {\enquote {\bibinfo {title} {{Extra U(1),
  effective operators, anomalies and dark matter}},}\ }\href {\doibase
  10.1007/JHEP11(2013)083} {\bibfield  {journal} {\bibinfo  {journal} {JHEP}\
  }\textbf {\bibinfo {volume} {11}},\ \bibinfo {pages} {083} (\bibinfo {year}
  {2013})},\ \Eprint {http://arxiv.org/abs/1307.0005} {arXiv:1307.0005
  [hep-ph]} \BibitemShut {NoStop}%
\bibitem [{\citenamefont {Ducu}\ \emph {et~al.}(2016)\citenamefont {Ducu},
  \citenamefont {Heurtier},\ and\ \citenamefont {Maurer}}]{Ducu:2015fda}%
  \BibitemOpen
  \bibfield  {author} {\bibinfo {author} {\bibfnamefont {Otilia}\ \bibnamefont
  {Ducu}}, \bibinfo {author} {\bibfnamefont {Lucien}\ \bibnamefont {Heurtier}},
  \ and\ \bibinfo {author} {\bibfnamefont {Julien}\ \bibnamefont {Maurer}},\
  }\bibfield  {title} {\enquote {\bibinfo {title} {{LHC signatures of a $Z'$
  mediator between dark matter and the SU(3) sector}},}\ }\href {\doibase
  10.1007/JHEP03(2016)006} {\bibfield  {journal} {\bibinfo  {journal} {JHEP}\
  }\textbf {\bibinfo {volume} {03}},\ \bibinfo {pages} {006} (\bibinfo {year}
  {2016})},\ \Eprint {http://arxiv.org/abs/1509.05615} {arXiv:1509.05615
  [hep-ph]} \BibitemShut {NoStop}%
\bibitem [{\citenamefont {Bhattacharyya}\ \emph {et~al.}(2018)\citenamefont
  {Bhattacharyya}, \citenamefont {Dutra}, \citenamefont {Mambrini},\ and\
  \citenamefont {Pierre}}]{Bhattacharyya:2018evo}%
  \BibitemOpen
  \bibfield  {author} {\bibinfo {author} {\bibfnamefont {Gautam}\ \bibnamefont
  {Bhattacharyya}}, \bibinfo {author} {\bibfnamefont {Ma\'ira}\ \bibnamefont
  {Dutra}}, \bibinfo {author} {\bibfnamefont {Yann}\ \bibnamefont {Mambrini}},
  \ and\ \bibinfo {author} {\bibfnamefont {Mathias}\ \bibnamefont {Pierre}},\
  }\bibfield  {title} {\enquote {\bibinfo {title} {{Freezing-in dark matter
  through a heavy invisible $Z'$}},}\ }\href {\doibase
  10.1103/PhysRevD.98.035038} {\bibfield  {journal} {\bibinfo  {journal} {Phys.
  Rev. D}\ }\textbf {\bibinfo {volume} {98}},\ \bibinfo {pages} {035038}
  (\bibinfo {year} {2018})},\ \Eprint {http://arxiv.org/abs/1806.00016}
  {arXiv:1806.00016 [hep-ph]} \BibitemShut {NoStop}%
\bibitem [{\citenamefont {Llorente}\ and\ \citenamefont
  {Nachman}(2018)}]{Llorente:2018wup}%
  \BibitemOpen
  \bibfield  {author} {\bibinfo {author} {\bibfnamefont {Javier}\ \bibnamefont
  {Llorente}}\ and\ \bibinfo {author} {\bibfnamefont {Benjamin~P.}\
  \bibnamefont {Nachman}},\ }\bibfield  {title} {\enquote {\bibinfo {title}
  {{Limits on new coloured fermions using precision jet data from the Large
  Hadron Collider}},}\ }\href {\doibase 10.1016/j.nuclphysb.2018.09.008}
  {\bibfield  {journal} {\bibinfo  {journal} {Nucl. Phys. B}\ }\textbf
  {\bibinfo {volume} {936}},\ \bibinfo {pages} {106--117} (\bibinfo {year}
  {2018})},\ \Eprint {http://arxiv.org/abs/1807.00894} {arXiv:1807.00894
  [hep-ph]} \BibitemShut {NoStop}%
\bibitem [{\citenamefont {Dutra}(2019)}]{Dutra:2019gqz}%
  \BibitemOpen
  \bibfield  {author} {\bibinfo {author} {\bibfnamefont {Maira}\ \bibnamefont
  {Dutra}},\ }\href@noop {} {Ph.D. thesis},\ \bibinfo  {school} {Orsay, LPT}
  (\bibinfo {year} {2019})\BibitemShut {NoStop}%
\bibitem [{\citenamefont {Kahlhoefer}\ \emph {et~al.}(2016)\citenamefont
  {Kahlhoefer}, \citenamefont {Schmidt-Hoberg}, \citenamefont {Schwetz},\ and\
  \citenamefont {Vogl}}]{Kahlhoefer:2015bea}%
  \BibitemOpen
  \bibfield  {author} {\bibinfo {author} {\bibfnamefont {Felix}\ \bibnamefont
  {Kahlhoefer}}, \bibinfo {author} {\bibfnamefont {Kai}\ \bibnamefont
  {Schmidt-Hoberg}}, \bibinfo {author} {\bibfnamefont {Thomas}\ \bibnamefont
  {Schwetz}}, \ and\ \bibinfo {author} {\bibfnamefont {Stefan}\ \bibnamefont
  {Vogl}},\ }\bibfield  {title} {\enquote {\bibinfo {title} {{Implications of
  unitarity and gauge invariance for simplified dark matter models}},}\ }\href
  {\doibase 10.1007/JHEP02(2016)016} {\bibfield  {journal} {\bibinfo  {journal}
  {JHEP}\ }\textbf {\bibinfo {volume} {02}},\ \bibinfo {pages} {016} (\bibinfo
  {year} {2016})},\ \Eprint {http://arxiv.org/abs/1510.02110} {arXiv:1510.02110
  [hep-ph]} \BibitemShut {NoStop}%
\bibitem [{\citenamefont {Alwall}\ \emph {et~al.}(2014)\citenamefont {Alwall},
  \citenamefont {Frederix}, \citenamefont {Frixione}, \citenamefont {Hirschi},
  \citenamefont {Maltoni}, \citenamefont {Mattelaer}, \citenamefont {Shao},
  \citenamefont {Stelzer}, \citenamefont {Torrielli},\ and\ \citenamefont
  {Zaro}}]{Alwall:2014hca}%
  \BibitemOpen
  \bibfield  {author} {\bibinfo {author} {\bibfnamefont {J.}~\bibnamefont
  {Alwall}}, \bibinfo {author} {\bibfnamefont {R.}~\bibnamefont {Frederix}},
  \bibinfo {author} {\bibfnamefont {S.}~\bibnamefont {Frixione}}, \bibinfo
  {author} {\bibfnamefont {V.}~\bibnamefont {Hirschi}}, \bibinfo {author}
  {\bibfnamefont {F.}~\bibnamefont {Maltoni}}, \bibinfo {author} {\bibfnamefont
  {O.}~\bibnamefont {Mattelaer}}, \bibinfo {author} {\bibfnamefont {H.~S.}\
  \bibnamefont {Shao}}, \bibinfo {author} {\bibfnamefont {T.}~\bibnamefont
  {Stelzer}}, \bibinfo {author} {\bibfnamefont {P.}~\bibnamefont {Torrielli}},
  \ and\ \bibinfo {author} {\bibfnamefont {M.}~\bibnamefont {Zaro}},\
  }\bibfield  {title} {\enquote {\bibinfo {title} {{The automated computation
  of tree-level and next-to-leading order differential cross sections, and
  their matching to parton shower simulations}},}\ }\href {\doibase
  10.1007/JHEP07(2014)079} {\bibfield  {journal} {\bibinfo  {journal} {JHEP}\
  }\textbf {\bibinfo {volume} {07}},\ \bibinfo {pages} {079} (\bibinfo {year}
  {2014})},\ \Eprint {http://arxiv.org/abs/1405.0301} {arXiv:1405.0301
  [hep-ph]} \BibitemShut {NoStop}%
\bibitem [{\citenamefont {Strigari}(2018)}]{Strigari:2018utn}%
  \BibitemOpen
  \bibfield  {author} {\bibinfo {author} {\bibfnamefont {Louis~E.}\
  \bibnamefont {Strigari}},\ }\bibfield  {title} {\enquote {\bibinfo {title}
  {{Dark matter in dwarf spheroidal galaxies and indirect detection: a
  review}},}\ }\href {\doibase 10.1088/1361-6633/aaae16} {\bibfield  {journal}
  {\bibinfo  {journal} {Rept. Prog. Phys.}\ }\textbf {\bibinfo {volume} {81}},\
  \bibinfo {pages} {056901} (\bibinfo {year} {2018})},\ \Eprint
  {http://arxiv.org/abs/1805.05883} {arXiv:1805.05883 [astro-ph.CO]}
  \BibitemShut {NoStop}%
\bibitem [{\citenamefont {Cirelli}\ \emph {et~al.}(2011)\citenamefont
  {Cirelli}, \citenamefont {Corcella}, \citenamefont {Hektor}, \citenamefont
  {Hutsi}, \citenamefont {Kadastik}, \citenamefont {Panci}, \citenamefont
  {Raidal}, \citenamefont {Sala},\ and\ \citenamefont
  {Strumia}}]{Cirelli:2010xx}%
  \BibitemOpen
  \bibfield  {author} {\bibinfo {author} {\bibfnamefont {Marco}\ \bibnamefont
  {Cirelli}}, \bibinfo {author} {\bibfnamefont {Gennaro}\ \bibnamefont
  {Corcella}}, \bibinfo {author} {\bibfnamefont {Andi}\ \bibnamefont {Hektor}},
  \bibinfo {author} {\bibfnamefont {Gert}\ \bibnamefont {Hutsi}}, \bibinfo
  {author} {\bibfnamefont {Mario}\ \bibnamefont {Kadastik}}, \bibinfo {author}
  {\bibfnamefont {Paolo}\ \bibnamefont {Panci}}, \bibinfo {author}
  {\bibfnamefont {Martti}\ \bibnamefont {Raidal}}, \bibinfo {author}
  {\bibfnamefont {Filippo}\ \bibnamefont {Sala}}, \ and\ \bibinfo {author}
  {\bibfnamefont {Alessandro}\ \bibnamefont {Strumia}},\ }\bibfield  {title}
  {\enquote {\bibinfo {title} {{PPPC 4 DM ID: A Poor Particle Physicist
  Cookbook for Dark Matter Indirect Detection}},}\ }\href {\doibase
  10.1088/1475-7516/2012/10/E01} {\bibfield  {journal} {\bibinfo  {journal}
  {JCAP}\ }\textbf {\bibinfo {volume} {03}},\ \bibinfo {pages} {051} (\bibinfo
  {year} {2011})},\ \bibinfo {note} {[Erratum: JCAP 10, E01 (2012)]},\ \Eprint
  {http://arxiv.org/abs/1012.4515} {arXiv:1012.4515 [hep-ph]} \BibitemShut
  {NoStop}%
\bibitem [{\citenamefont {Ambrogi}\ \emph {et~al.}(2019)\citenamefont
  {Ambrogi}, \citenamefont {Arina}, \citenamefont {Backovic}, \citenamefont
  {Heisig}, \citenamefont {Maltoni}, \citenamefont {Mantani}, \citenamefont
  {Mattelaer},\ and\ \citenamefont {Mohlabeng}}]{Ambrogi:2018jqj}%
  \BibitemOpen
  \bibfield  {author} {\bibinfo {author} {\bibfnamefont {Federico}\
  \bibnamefont {Ambrogi}}, \bibinfo {author} {\bibfnamefont {Chiara}\
  \bibnamefont {Arina}}, \bibinfo {author} {\bibfnamefont {Mihailo}\
  \bibnamefont {Backovic}}, \bibinfo {author} {\bibfnamefont {Jan}\
  \bibnamefont {Heisig}}, \bibinfo {author} {\bibfnamefont {Fabio}\
  \bibnamefont {Maltoni}}, \bibinfo {author} {\bibfnamefont {Luca}\
  \bibnamefont {Mantani}}, \bibinfo {author} {\bibfnamefont {Olivier}\
  \bibnamefont {Mattelaer}}, \ and\ \bibinfo {author} {\bibfnamefont
  {Gopolang}\ \bibnamefont {Mohlabeng}},\ }\bibfield  {title} {\enquote
  {\bibinfo {title} {{MadDM v.3.0: a Comprehensive Tool for Dark Matter
  Studies}},}\ }\href {\doibase 10.1016/j.dark.2018.11.009} {\bibfield
  {journal} {\bibinfo  {journal} {Phys. Dark Univ.}\ }\textbf {\bibinfo
  {volume} {24}},\ \bibinfo {pages} {100249} (\bibinfo {year} {2019})},\
  \Eprint {http://arxiv.org/abs/1804.00044} {arXiv:1804.00044 [hep-ph]}
  \BibitemShut {NoStop}%
\bibitem [{\citenamefont {Hannestad}(2004)}]{Hannestad:2004px}%
  \BibitemOpen
  \bibfield  {author} {\bibinfo {author} {\bibfnamefont {Steen}\ \bibnamefont
  {Hannestad}},\ }\bibfield  {title} {\enquote {\bibinfo {title} {{What is the
  lowest possible reheating temperature?}}}\ }\href {\doibase
  10.1103/PhysRevD.70.043506} {\bibfield  {journal} {\bibinfo  {journal} {Phys.
  Rev. D}\ }\textbf {\bibinfo {volume} {70}},\ \bibinfo {pages} {043506}
  (\bibinfo {year} {2004})},\ \Eprint {http://arxiv.org/abs/astro-ph/0403291}
  {arXiv:astro-ph/0403291} \BibitemShut {NoStop}%
\bibitem [{\citenamefont {Jedamzik}(2006)}]{Jedamzik:2006xz}%
  \BibitemOpen
  \bibfield  {author} {\bibinfo {author} {\bibfnamefont {Karsten}\ \bibnamefont
  {Jedamzik}},\ }\bibfield  {title} {\enquote {\bibinfo {title} {{Big bang
  nucleosynthesis constraints on hadronically and electromagnetically decaying
  relic neutral particles}},}\ }\href {\doibase 10.1103/PhysRevD.74.103509}
  {\bibfield  {journal} {\bibinfo  {journal} {Phys. Rev. D}\ }\textbf {\bibinfo
  {volume} {74}},\ \bibinfo {pages} {103509} (\bibinfo {year} {2006})},\
  \Eprint {http://arxiv.org/abs/hep-ph/0604251} {arXiv:hep-ph/0604251}
  \BibitemShut {NoStop}%
\bibitem [{\citenamefont {Kawasaki}\ \emph {et~al.}(2018)\citenamefont
  {Kawasaki}, \citenamefont {Kohri}, \citenamefont {Moroi},\ and\ \citenamefont
  {Takaesu}}]{Kawasaki:2017bqm}%
  \BibitemOpen
  \bibfield  {author} {\bibinfo {author} {\bibfnamefont {Masahiro}\
  \bibnamefont {Kawasaki}}, \bibinfo {author} {\bibfnamefont {Kazunori}\
  \bibnamefont {Kohri}}, \bibinfo {author} {\bibfnamefont {Takeo}\ \bibnamefont
  {Moroi}}, \ and\ \bibinfo {author} {\bibfnamefont {Yoshitaro}\ \bibnamefont
  {Takaesu}},\ }\bibfield  {title} {\enquote {\bibinfo {title} {{Revisiting
  Big-Bang Nucleosynthesis Constraints on Long-Lived Decaying Particles}},}\
  }\href {\doibase 10.1103/PhysRevD.97.023502} {\bibfield  {journal} {\bibinfo
  {journal} {Phys. Rev. D}\ }\textbf {\bibinfo {volume} {97}},\ \bibinfo
  {pages} {023502} (\bibinfo {year} {2018})},\ \Eprint
  {http://arxiv.org/abs/1709.01211} {arXiv:1709.01211 [hep-ph]} \BibitemShut
  {NoStop}%
\bibitem [{\citenamefont {Giudice}\ \emph {et~al.}(2001)\citenamefont
  {Giudice}, \citenamefont {Kolb},\ and\ \citenamefont
  {Riotto}}]{Giudice:2000ex}%
  \BibitemOpen
  \bibfield  {author} {\bibinfo {author} {\bibfnamefont {Gian~Francesco}\
  \bibnamefont {Giudice}}, \bibinfo {author} {\bibfnamefont {Edward~W.}\
  \bibnamefont {Kolb}}, \ and\ \bibinfo {author} {\bibfnamefont {Antonio}\
  \bibnamefont {Riotto}},\ }\bibfield  {title} {\enquote {\bibinfo {title}
  {{Largest temperature of the radiation era and its cosmological
  implications}},}\ }\href {\doibase 10.1103/PhysRevD.64.023508} {\bibfield
  {journal} {\bibinfo  {journal} {Phys. Rev. D}\ }\textbf {\bibinfo {volume}
  {64}},\ \bibinfo {pages} {023508} (\bibinfo {year} {2001})},\ \Eprint
  {http://arxiv.org/abs/hep-ph/0005123} {arXiv:hep-ph/0005123} \BibitemShut
  {NoStop}%
\bibitem [{\citenamefont {Baum}\ \emph {et~al.}(2018)\citenamefont {Baum},
  \citenamefont {Catena}, \citenamefont {Conrad}, \citenamefont {Freese},\ and\
  \citenamefont {Krauss}}]{Baum:2017kfa}%
  \BibitemOpen
  \bibfield  {author} {\bibinfo {author} {\bibfnamefont {Sebastian}\
  \bibnamefont {Baum}}, \bibinfo {author} {\bibfnamefont {Riccardo}\
  \bibnamefont {Catena}}, \bibinfo {author} {\bibfnamefont {Jan}\ \bibnamefont
  {Conrad}}, \bibinfo {author} {\bibfnamefont {Katherine}\ \bibnamefont
  {Freese}}, \ and\ \bibinfo {author} {\bibfnamefont {Martin~B.}\ \bibnamefont
  {Krauss}},\ }\bibfield  {title} {\enquote {\bibinfo {title} {{Determining
  dark matter properties with a XENONnT/LZ signal and LHC Run 3 monojet
  searches}},}\ }\href {\doibase 10.1103/PhysRevD.97.083002} {\bibfield
  {journal} {\bibinfo  {journal} {Phys. Rev. D}\ }\textbf {\bibinfo {volume}
  {97}},\ \bibinfo {pages} {083002} (\bibinfo {year} {2018})},\ \Eprint
  {http://arxiv.org/abs/1709.06051} {arXiv:1709.06051 [hep-ph]} \BibitemShut
  {NoStop}%
\bibitem [{\citenamefont {{ATLAS
  Collaboration}}(2022)}]{ATL-PHYS-PUB-2022-018}%
  \BibitemOpen
  \bibfield  {author} {\bibinfo {author} {\bibnamefont {{ATLAS
  Collaboration}}},\ }\href {http://cds.cern.ch/record/2805993} {\enquote
  {\bibinfo {title} {{Snowmass White Paper Contribution: Physics with the
  Phase-2 ATLAS and CMS Detectors}},}\ } (\bibinfo {year} {2022})\BibitemShut
  {NoStop}%
\bibitem [{\citenamefont {Viana}\ \emph {et~al.}(2021)\citenamefont {Viana},
  \citenamefont {Albert}, \citenamefont {Harding}, \citenamefont {Hinton},
  \citenamefont {Schoorlemmer},\ and\ \citenamefont
  {de~Souza}}]{Viana:2021smp}%
  \BibitemOpen
  \bibfield  {author} {\bibinfo {author} {\bibfnamefont {Aion}\ \bibnamefont
  {Viana}}, \bibinfo {author} {\bibfnamefont {Andrea}\ \bibnamefont {Albert}},
  \bibinfo {author} {\bibfnamefont {J.~Patrick}\ \bibnamefont {Harding}},
  \bibinfo {author} {\bibfnamefont {Jim}\ \bibnamefont {Hinton}}, \bibinfo
  {author} {\bibfnamefont {Harm}\ \bibnamefont {Schoorlemmer}}, \ and\ \bibinfo
  {author} {\bibfnamefont {Vitor}\ \bibnamefont {de~Souza}} (\bibinfo
  {collaboration} {SWGO}),\ }\bibfield  {title} {\enquote {\bibinfo {title}
  {{Searching for Dark Matter with the Southern Wide-field Gamma-ray
  Observatory (SWGO)}},}\ }\href {\doibase 10.22323/1.395.0555} {\bibfield
  {journal} {\bibinfo  {journal} {PoS}\ }\textbf {\bibinfo {volume}
  {ICRC2021}},\ \bibinfo {pages} {555} (\bibinfo {year} {2021})}\BibitemShut
  {NoStop}%
\bibitem [{\citenamefont {Delos}\ and\ \citenamefont
  {Linden}(2022)}]{Delos:2021rqs}%
  \BibitemOpen
  \bibfield  {author} {\bibinfo {author} {\bibfnamefont {M.~S.}\ \bibnamefont
  {Delos}}\ and\ \bibinfo {author} {\bibfnamefont {Tim}\ \bibnamefont
  {Linden}},\ }\bibfield  {title} {\enquote {\bibinfo {title} {{Dark matter
  microhalos in the solar neighborhood: Pulsar timing signatures of early
  matter domination}},}\ }\href {\doibase 10.1103/PhysRevD.105.123514}
  {\bibfield  {journal} {\bibinfo  {journal} {Phys. Rev. D}\ }\textbf {\bibinfo
  {volume} {105}},\ \bibinfo {pages} {123514} (\bibinfo {year} {2022})},\
  \Eprint {http://arxiv.org/abs/2109.03240} {arXiv:2109.03240 [astro-ph.CO]}
  \BibitemShut {NoStop}%
\end{thebibliography}

%

\end{document}